\documentclass[journal]{IEEEtran}

\usepackage{amsmath,amssymb,amsfonts}
\usepackage{graphicx}
\usepackage{xcolor}
\usepackage{hyperref}
\usepackage[labelformat=simple]{subcaption}

\usepackage{tikz}
\usepackage{makecell}
\usepackage[utf8]{inputenc}
\usepackage{multirow}

\DeclareMathOperator{\dB}{dB}
\DeclareMathOperator{\PL}{PL}
\DeclareMathOperator{\power}{P}
\DeclareMathOperator{\RM}{RM}

\DeclareMathOperator{\dBm}{dBm}
\DeclareMathOperator{\watts}{W}
\DeclareMathOperator{\SINR}{SINR}
\DeclareMathOperator{\Hz}{Hz}
\DeclareMathOperator{\MHz}{MHz}
\DeclareMathOperator{\SNR}{SNR}
\DeclareMathOperator{\trnc}{trnc}
\DeclareMathOperator{\thr}{thr}
\DeclareMathOperator{\Tx}{T}
\DeclareMathOperator{\Rx}{R}

\newcommand*{\jim}{\mathfrak{j}}
\newcommand*{\expec}{\mathbb{E}}
\newcommand*{\dif}{\,\text{d}}
\newcommand*{\NoiseFloor}{\mathcal{N}}
\newcommand*{\NoiseFigure}{\text{NF}}

\title{Radio Map Prediction from Aerial Images and Application to Coverage Optimization}

\author{Fabian Jaensch, 
Giuseppe Caire \IEEEmembership{Fellow, IEEE},
Beg{\"u}m Demir \IEEEmembership{Senior Member, IEEE}
\thanks{F. Jaensch and G. Caire are with the Communications and Information Theory Group, Technische Universit{\"a}t Berlin, 10623 Berlin, Germany.}
\thanks{B. Demir is with the Remote Sensing Image Analysis Group, Technische Universit{\"a}t Berlin, 10623 Berlin, Germany and with the BIFOLD - Berlin Institute 
for the Foundations of Learning and Data, 10623 Berlin, Germany.}
\thanks{Corresponding author: Fabian Jaensch (email: f.jaensch@tu-berlin.de).}
\thanks{The work of G. Caire and F. Jaensch was supported by BMBF Germany in the program of ``Souverän. Digital. Vernetzt.'' Joint Project 6G-RIC (Project IDs 16KISK030).}
\thanks{To appear in IEEE Transactions on Wireless Communications © 2025 IEEE.  Personal use of this material is permitted.  Permission from IEEE must be obtained for all other uses, in any current or future media, including reprinting/republishing this material for advertising or promotional purposes, 
creating new collective works, for resale or redistribution to servers or lists, or reuse of any copyrighted component of this work in other works.}
}

\begin{document}
    \maketitle
    
    \begin{abstract}
        Several studies have explored deep learning algorithms to predict large-scale signal fading, or path loss, in urban communication networks. 
        The goal is to replace costly measurement campaigns, inaccurate statistical models, or computationally expensive ray-tracing simulations with machine learning models that deliver quick and accurate predictions.

        We focus on predicting path loss radio maps using convolutional neural networks, leveraging aerial images alone or in combination with supplementary height information. 
        Notably, our approach does not rely on explicit classification of environmental objects, which is often unavailable for most locations worldwide. 
        While the prediction of radio maps using complete 3D environmental data is well-studied, the use of only aerial images remains under-explored. 
        We address this gap by showing that state-of-the-art models developed for existing radio map datasets can be effectively adapted to this task. 
        Additionally, we introduce a new model dubbed UNetDCN that achieves on par or better performance compared to the state-of-the-art with reduced complexity.

        The trained models are differentiable, and therefore they can be incorporated in various network optimization algorithms. 
        While an extensive discussion is beyond this paper's scope, we demonstrate this through an example optimizing the directivity of base stations in cellular networks via backpropagation to enhance coverage.
    \end{abstract}

    \begin{IEEEkeywords}
        Convolutional Neural Networks, Machine Learning, Path loss, Radio map, RSSI, Coverage
    \end{IEEEkeywords}

    \section{INTRODUCTION}\label{sec:introduction}
        \IEEEPARstart{W}{ireless}
        communication systems are fundamentally based on radio waves radiated by a transmitter (Tx) antenna carrying a signal to send information to a receiving (Rx) antenna.
        To model the effects of the propagation channel between Tx and Rx, it is convenient to distinguish between ``small-scale'' and ``large-scale'' effects. 
        The small-scale effects are due to the constructive/destructive interference between the propagation paths that add with different phases, amplitudes, and delays at the Rx antenna. 
        This results in a time and frequency selective fading process usually modeled as a correlated Gaussian random process with normalized second moment. 
        In contrast, the large-scale effects capture the attenuation of the received signal power, or path loss.  
        This attenuation depends on distance-dependent power dissipation in free space, reflections from building and street surfaces, and attenuation by tree canopies. 
        To capture this quantity, we consider the radio map of a transmitter consisting of the path loss values for different potential Rx locations of interest.

        Important applications of radio maps include the optimization of cellular network layout to achieve good coverage of an area (Section \ref{sec:optimization}), link scheduling \cite{link_scheduling}, localization of user equipment (UE) based on RSS measurements at the Rx \cite{locunet} or, in the context of 5G and 6G, beam management \cite{beams}.
        Since measurement campaigns are impractical and too expensive on a large scale, different methods have been developed to approximate the path loss.
        Statistical path loss models that express the attenuation solely as a function of the distance between Tx and Rx and additional stochastic effects are inherently unable to capture the radio wave propagation in a specific environment.
        Ray-tracing simulations provide an option to approximate the underlying physical phenomena with high accuracy \cite{ray-tracing-accuracy}. 
        However, due to the relatively long run times of the simulations and the need for detailed 3D city models, they can be infeasible for applications in real time or on large scale.

        Recently, several works have explored the application of machine learning (ML) algorithms and, in particular, deep neural networks (DNNs) to the path loss prediction problem.
        The general idea is that a properly designed ML model can learn the underlying physical phenomena of radio wave propagation to a certain extent, given a sufficient amount of training data.
        Albeit the training process can be very time- and resource-intensive, the run time of the inference task once the model is deployed is typically in the order of milliseconds.

        This opens up a range of applications in which statistical path loss models are too inaccurate and ray-tracing is too slow, \cite{pmnet}, \cite{applications}. 
        Additionally, DNNs are a very flexible tool for extracting information and patterns implicitly from data.
        We show that convolutional neural networks (CNN) are capable of implicitly extracting information about the presence, shape, and height of buildings and trees that is necessary to determine radio maps from aerial images.

        In the remainder of this Section, we give an account of the relevant literature for radio map prediction, in particular using deep learning, and a comparison to our contributions. 
        Section \ref{sec:system_methodology} reviews some necessary wireless communication background to precisely define the investigated problem, followed by a description of the generation of the dataset, CNN architectures and other aspects of the experiments.
        In Section \ref{sec:numerical_results}, we discuss the results on radio map prediction and provide an ablation study for the proposed UNetDCN.
        Lastly, Section \ref{sec:optimization} showcases applications to network coverage optimization and Section \ref{sec:conclusions} finishes with the conclusions.

    \subsection{General Approach and Related Works}\label{sec:related_works}
        A common way to model the path loss is to use the log normal shadowing models in different variations (e.g. the 3GPP model described in \cite{3gpp_log_normal}).
        They originate from the idea to express the path loss merely as a function of the distance between Tx and Rx and to model deviations as the realization of a Gaussian random variable in the log (dB) domain.
        Albeit the resulting graph of the function together with the stochastic spread may fit measurement points, these models completely disregard the radio wave propagation in a specific environment along specific paths.

        Ray-tracing \cite{ray-tracing} provides a more accurate, site-specific solution by modeling electromagnetic waves as rays launched in a very fine subdivision of the space around the Tx.
        Reflections from surfaces and potentially other effects such as diffractions or transmissions are calculated upon contact with objects \cite{ray-tracing-accuracy}.
        The rays arriving at the Rx are combined to generate the path loss as defined in Section \ref{sec:pathloss}, \eqref{eq:power_ratio}, or other channel state information.
        As mentioned before, potential drawbacks for certain applications are the runtime of typically at least several seconds to generate a complete radio map and the need to have a complete 3D model of the environment available.

        Other methods include completing the radio map from sparse measurements \cite{rm_completion} or estimating the path loss for a single Rx position at a time via ML \cite{ml_single}.

        In the following, we describe approaches to predict the whole radio map at once using ML.
        Each pixel in the target radio map, represented by a two-dimensional image, corresponds to a different location in the considered environment from a birds-eye perspective, and its value is the path loss on an appropriate scale.
        The input information about the presence and potentially height of objects like buildings or trees, the transmitter position, and possibly other parameters that differ from sample to sample are usually encoded in two-dimensional images of the same shape as the target and stacked along an additional channel dimension.
        All works described in the following build on CNN models \cite{cnn}, which have been shown to be effective in other tasks such as semantic segmentation, image denoising, or image-to-image translation that are structurally similar in the sense that input and output of the model are image-like tensors. 

        A first seminal contribution investigating the prediction of the whole radio map at once using CNNs came out of our research group \cite{radiounet} and considered a UNet \cite{unet} model trained in a supervised manner to approximate radio maps generated with two-dimensional ray-tracing from city maps and Tx locations.
        The presence or absence of buildings and the Tx are encoded in binary input images.
        The authors explore strategies to incorporate available signal measurements and adapt the trained models to more realistic scenarios and show applications to coverage estimation and fingerprint localization.
        Several other works have extended this approach to the more challenging scenario of radio maps generated with 3D ray-tracing, e.g. \cite{plnet, fadenet, radiotrans}.
        In these works, the binary input information per pixel is usually replaced by height information.
        Some works have  experimented with more variables between the different samples encoded in additional inputs, for example the carrier frequency of the transmitted signal \cite{radiotrans} or antenna patterns and orientations \cite{plnet}.
        Albeit the basic encoder-decoder with skip-connections design of the UNet has been used in most works since then, it has been soon recognized that standard UNets are inherently limited when it comes to propagating information over longer distances, which is especially important to accurately predict reflections.
        Several approaches make use of dilated convolutions \cite{dilated} to increase the receptive field \cite{ziemann21,deepray,pmnet}, whereas in \cite{radiotrans} vision transformer layers \cite{vit} are added to the CNN to allow modeling long-range relationships.
        Besides changes to the network architecture, some authors propose feature engineering to resolve the described problem and improve the accuracy.
        In \cite{qiu22}, the network is provided with an input image containing the spatial distance of each pixel position to the transmitter location.
        The authors of \cite{radiotrans} propose to feed the coordinates of each pixel and the Tx to the network in all positions.
        To show the validity of training CNNs on simulated data, in \cite{plnet} a model tested first on radio maps generated by ray-tracing is retrained on real-world measurements and shown to perform better than all conventional methods.

        Only one work we are aware of investigates path loss prediction for a whole area from aerial images \cite{plgan}.
        The scenario considered in that study differs from ours as Tx are mounted on unmanned aerial vehicles (UAV) instead of buildings, and the developed model seems rather inaccurate with RMSE over $43\dB$.

        The authors of \cite{radiounet} and \cite{pmnet} regard general application of radio maps to coverage area classification and UE localization, which are not specifically requiring DNNs.
        In \cite{fadenet} it is shown that by making use of the fast inference speed of the trained model, a large number of combinations of potential Tx locations can be evaluated quickly in terms of the coverage provided in an area,
        reducing the time needed to find the optimal combination significantly compared to using ray-tracing.
        A review on more applications is provided in \cite{applications}.

        \begin{table*}
            \small
            \begin{tabular}{|c||c|c|c|c|}
                \hline
                Dataset         &   RadioMapSeer \cite{radiounet} &   RadioMapSeer3D \cite{cagkan3d} & USC \cite{pmnet} & RMDirectionalBerlin (ours) \\
                \hline
                \hline
                Dimension       &   2D      &       3D      &       2D      &   3D  \\
                \hline
                Environment data&   buildings &   \makecell[tc]{buildings with a single\\ random height value}  &   buildings &   building and vegetation nDSMs   \\            
                \hline
                Antennas (Tx)   &   isotropic   &   isotropic   &   isotropic   &   directional \\
                \hline
                \#Samples       &   $56080$ &  $50680$  &   $19016$ &   $74515$     \\
                \hline
                Additional data &   \makecell[tc]{repeated simulations with\\different propagation models\\and additional cars,\\time-of-arrival maps} & -  &   -   &   \makecell[tc]{aerial images}\\
                \hline
            \end{tabular}
            \caption{Overview of open radio map datasets.}
            \label{table:datasets}
        \end{table*}

    \subsection{Our Contribution}\label{sec:contribution}
        We investigate the prediction of radio maps from aerial images and potentially unclassified height maps for scenarios in which a complete 3D model of the environment is not available.
        The radio map is only determined from the information about the environment and the Tx.
        In particular, contrary to some other works, we do not assume measurements to be available.
        Our results (RMSE of $6.7$dB without height information and $5.2$dB with height information) show a significant improvement over the only other work we are aware of considering a similar scenario ($44$dB in \cite{plgan}).

        For this purpose, we have generated city geometries from real places in the city of Berlin and conducted ray-tracing simulations to obtain a large collection of path loss maps modeling typical urban cellular networks, described in Section \ref{sec:dataset}.
        In contrast to the other publicly available radio map datasets we are aware of \cite{radiounet,pmnet,cagkan3d}, ours is the first one to feature directional antennas at the Tx and to include trees and realistic building heights with approximated roof shapes taken from the real world as well as aerial images.
        Table \ref{table:datasets} provides an overview of the similarities and differences between other openly accessible datasets and ours.
        \footnote{The dataset and some numerical comparisons of model architectures and input features have already been presented in our preprint \cite{jaensch2024radiomapestimation}. 
        For this paper, we have focused more deeply on the investigation of prediction from images, refined our model architecture, and added the investigation of applications to network coverage optimization.
        Our dataset can be found at \url{https://zenodo.org/records/13834313}, the code at \url{https://github.com/fabja19/RML_v2_img}.}

        We perform experiments with different CNN architectures and input features in order to encode the city geometry, the relative position of the Tx to locations on the map, and the Tx antenna characteristics, including several ideas proposed in the literature (Section \ref{sec:input_features}).
        Lastly, we consider the usage of deformable convolutional layers \cite{dcn} for the radio map prediction task and show that our proposed model achieves the same or higher accuracy as the state of the art, depending on the scenario, with reduced complexity.

        In Section \ref{sec:optimization}, we show how a trained model can be used to solve an optimization problem in a cellular network.
        In contrast to applications in previous works, such as \cite{fadenet}, where the fast inference speed of the trained network is used to optimize BS deployment via exhaustive search over all possible combinations of locations, we also take advantage of backpropagation through a trained and frozen network.
        This allows us to directly optimize input parameters defined as trainable variables with respect to some loss or score function calculated from the predicted radio maps via gradient descent.
        Although this technique has been used in e.g. inverse material design \cite{nanophotonics_inverse_design}, we are not aware of previous usage in our field.
        Note that the optimization of antenna characteristics and materials is also possible with the recently developed ray-tracer Sionna \cite{sionna}.
        Our approach, however, does not require a 3D model of the environment and works just with aerial images as inputs, whereas Sionna requires  an explicit 3D model as input.
        Additionally, optimization using our models is significantly faster than with Sionna, which has to recalculate the radio map via ray-tracing in each iteration \cite{sionna_rt}.

    \section{SYSTEM AND METHODOLOGY}\label{sec:system_methodology}
    \subsection{Wireless Communication Background}\label{sec:pathloss}
            Consider a Tx-Rx pair in a fixed environment with $n$ paths establishing the link from the Tx to the Rx.
            These may include a direct line-of-sight path and multipaths undergoing reflections from surfaces such as the ground or building walls and diffractions around edges or corners of objects.
            In our simulation scenario described in Section \ref{sec:dataset}, we assume that the wavelength of the communication signal is significantly smaller than the accuracy of spatial positions and shapes of objects in the environment.
            This implies that it is impossible to model the phase differences between different paths appropriately, and any small-scale fading due to constructive and destructive interference between paths arriving with different phase shifts is modeled as a random effect.
            In fact, in standard wireless communication theory it is customary to separate the large-scale path loss from the small-scale random fading and consider normalized statistics for the latter (e.g., Rayleigh or Rician fading with unit second moment, see \cite{molisch}). 

            We consider a time-varying uncorrelated scattering model \cite{wssus} with transfer function
            \begin{equation*}
                H(t, f) =   \sum_{p=1}^n c_p e^{\jim\psi_p} e^{\jim 2\pi\nu_p t} e^{-\jim 2\pi f\tau_p},
            \end{equation*}
            where $c_p\in\mathbb{C}$ is the complex amplitude of the $p$-th path coefficient, $\psi_p$ the corresponding phase factor that depends on the radio between the path length $d_p$ and the carrier wavelength $\lambda$, $\nu_p$ is the Doppler shift, $\tau_p$ the delay, $t,f$ denote time and frequency, respectively, and $\jim$ is the imaginary unit.
            We define the ratio between received and transmitted power, i.e., the channel power attenuation or \textit{path loss}, as
            \begin{equation}\label{eq:power_ratio}
                \frac{\power_{\Rx}}{\power_{\Tx}} = \sum_{p=1}^n \expec\left[|c_p|^2\right],
            \end{equation}
            in order to capture the large-scale fading and average out random fluctuations. 
            More precisely, since the coefficients $c_p$ are the result of a large number of microscopic multipath components adding with slightly different phases, they are modeled as complex circularly symmetric random variables
            \begin{equation*}
                c_p \sim \mathcal{CN}(\mu_p, \sigma^2_p) 
            \end{equation*}
            such that $\expec[|c_p|^2] = \left|\mu_p\right|^2 + \sigma_p^2$ is the power attenuation along the $p$-th path.  
            In addition, since the phase $\psi_p$ depends on $\frac{d_p}{\lambda}$, and since $\frac{2 \pi d_p}{\lambda} \mod{2 \pi}$ can be any number in $[0, 2\pi]$ depending on the very exact details of the environment, it is reasonable to model these phases as mutually independent, independent of the $c_p$, and uniformly distributed in $[0,2\pi]$. 
            Therefore, $\phi_p - \phi_q$ is uniformly distributed over $[0, 2\pi]$ for $p \neq q$ and it follows that 
            \begin{equation}\label{eq:rss_expected_g}
                \begin{split}
                    &\expec[ |H(t, f)|^2] \\
                        = &\sum_{p,q=1}^n \expec[ c_p c_q^* e^{\jim(c_p - c_q)}] e^{\jim2\pi(\nu_p - \nu_q)t} e^{-\jim2\pi (\tau_p - \tau_q)f} \\
                        = &\sum_{p=1}^n \expec[|c_p|^2],
                \end{split}
            \end{equation}
            where $\cdot^*$ denotes the complex conjugate, thus justifying definition \eqref{eq:power_ratio} for the large-scale path loss. 

            We notice here that the ensemble expectation in \eqref{eq:rss_expected_g} can also be justified in terms of local averaging over time and frequency of the instantaneous received signal power. 
            In fact, in practice, the received signal power is always measured at the receiver as a local average. Even if the coefficients $c_p e^{\jim\phi_p}$ are not modeled as random variables, but are completely deterministic, considering a time window of duration $T$ and the channel of bandwidth $W$, we have
            \begin{equation*}
                \begin{split}
                    \frac{\power_{\Rx}}{\power_{\Tx}} =& \frac{1}{W T} \int_0^T \int_{-W/2}^{W/2} \left|H(t, f)\right|^2 \dif t \dif f \\
                    =& \frac{1}{W T} \int_0^T \int_{-W/2}^{W/2}  \sum_{p,q=1}^n  c_p c_q^* e^{j (\phi_p - \phi_q)} e^{\jim 2\pi (\nu_p - \nu_q)t} \\
                    & e^{-\jim 2\pi (\tau_p - \tau_q) f} \dif t \dif f \\ 
                    \approx& \sum_{p=1}^n |c_p|^2 
                \end{split}
            \end{equation*}
            since 
            \begin{equation*}
                \frac{1}{T} \int_0^T e^{j 2\pi (\nu_p - \nu_q)t} \dif t  \approx 0  
            \end{equation*}

            for all $\nu_p \neq \nu_q $ when $|\nu_p - \nu_q|  > 1/T$, i.e., when the time variations due to the Doppler spread are significant over the averaging interval, and 

            \begin{equation*}
                \frac{1}{W} \int_{-W/2}^{W/2} e^{- \jim 2\pi (\tau_p - \tau_q) f} \dif f \approx 0   
            \end{equation*}

            for all $\tau_p \neq \tau_q $
            when $|\tau_p - \tau_q| > 1/W$, i.e., when the frequency variations due to the delay spread are significant over the channel bandwidth. 
            Note also that if two paths have both Doppler shift difference less than $1/T$ and delay difference less than $1/W$, then they can be counted as a single path since they are indistinguishable \cite{proakis}.
            Hence, the above conditions are always satisfied for all distinguishable paths.
            We conclude that whether we model the path coefficients as uncorrelated random variables and adopt the definition of average power as an ensemble average, or model them as deterministic quantities and use the definition of average power as a local time-frequency average, the macroscopic path loss is consistently defined as the sum of the squares (or the sum of the second moments) of the path coefficients.  

            The \textit{radio map} of the Tx for a fixed set $\mathbb{D}\subset\mathbb{R}^2$ of locations of interest may now formally be defined as a function $\RM:\mathbb{D}\rightarrow\mathbb{R}$, which maps each location to the path loss value for an Rx in the considered position according to \eqref{eq:power_ratio}.
            In the following, we will only consider the case that $\mathbb{D}$ corresponds to a uniform grid over a square-shaped area.
            This allows us to regard the radio map as a matrix or image. 

            As elaborated in \cite{radiounet}, it is reasonable to truncate the path loss from below at a threshold corresponding to the noise floor, as lower power levels are irrelevant in practice, and to work with powers and losses in $\dB$ scale, see Section \ref{sec:rm_simulations}. 
            
            \subsection{Dataset}\label{sec:dataset}
            Our dataset consists of 74,515 radio maps simulated using ray-tracing on 424 city maps from Berlin.
            The environment data and the radio maps cover an area of $256$m$\times256$m with a spatial resolution of $1$m.
            For each city map, several potential Tx locations on the corners and edges of building rooftops were identified. Simulations were conducted for each location using different antenna characteristics and orientations.
            \begin{figure}[!htbp]
                \centering
                \begin{subfigure}[t]{0.1\textwidth}
                    \includegraphics[width=\textwidth]{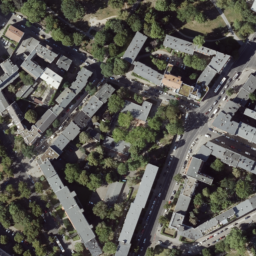}
                    \caption{}
                    \label{fig:maps:img}
                \end{subfigure}
                \begin{subfigure}[t]{0.1\textwidth}
                    \includegraphics[width=\textwidth]{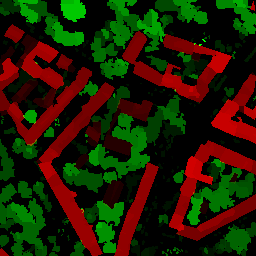}
                    \caption{}
                    \label{fig:maps:gis}
                \end{subfigure}
                \begin{subfigure}[t]{0.125\textwidth}
                    \includegraphics[width=\textwidth]{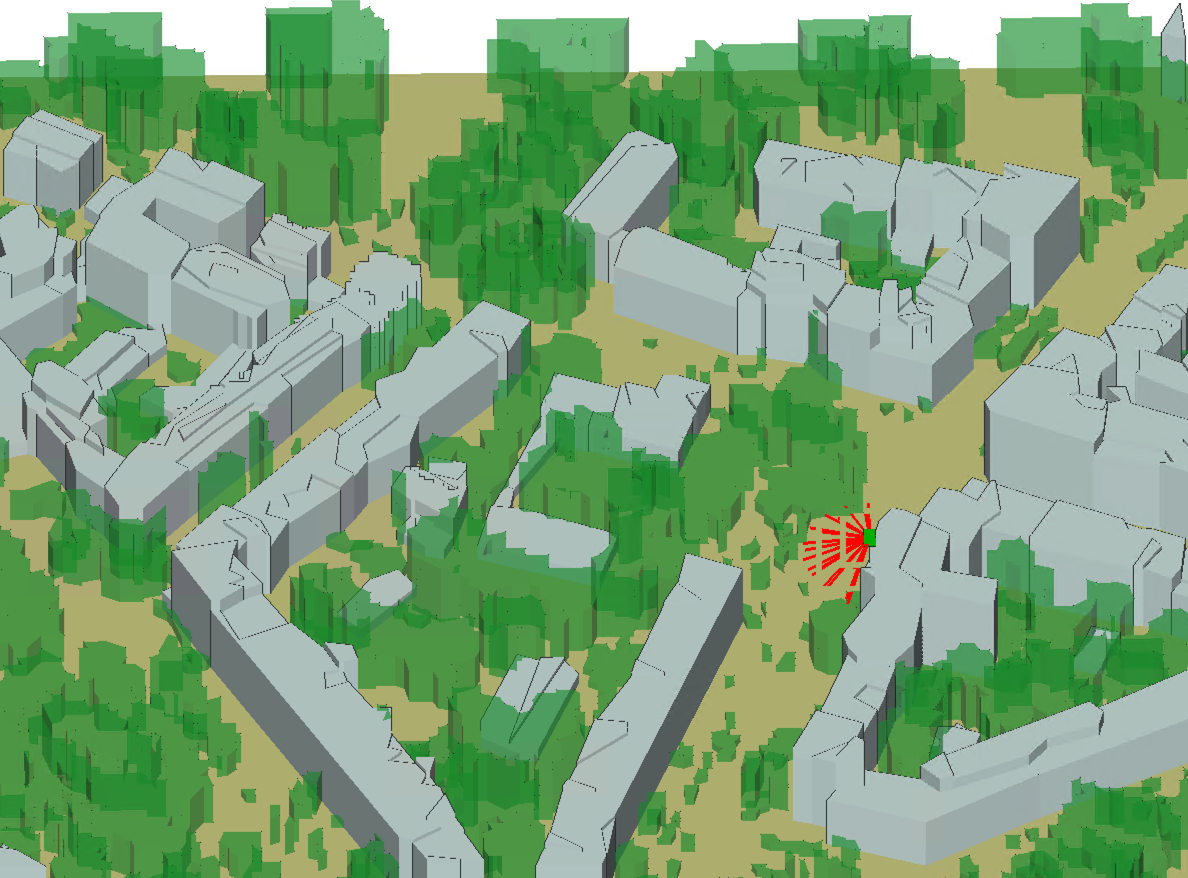}
                    \caption{}
                    \label{fig:maps:3d}
                \end{subfigure}
                \begin{subfigure}[t]{0.1\textwidth}
                    \includegraphics[width=\textwidth]{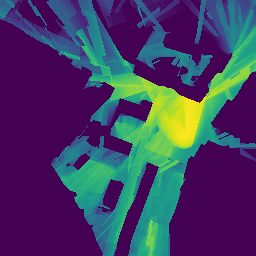}
                    \caption{}
                    \label{fig:maps:rm}
                \end{subfigure}
                \begin{subfigure}[t]{0.027\textwidth}
                    \includegraphics[width=\textwidth]{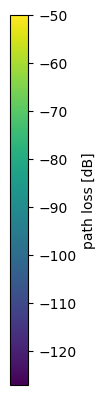}
                \end{subfigure}
                \caption{Sample from the dataset - \subref{fig:maps:img} aerial image \subref{fig:maps:gis} nDSMs of buildings (red) and vegetation (green) overlaid, with darker colors corresponding to greater height values \subref{fig:maps:3d}
                3D model in simulation with Tx (green cube) and different orientations of the antenna main lobe (red lines) \subref{fig:maps:rm} example of a radio map}
                \label{fig:maps}
            \end{figure}
        \subsubsection{City Maps}\label{sec:city_maps}
        Our goal was to simulate cellular networks in urban environments corresponding to places in the real world and featuring buildings  with realistic shapes and heights and also trees.
        Although a few public institutions provide open 3D models of certain cities, municipalities, or even whole countries, we could not find any that contained information about foliage.
        However, available height maps or digital surface models (DSMs) lack the necessary classification of objects into buildings, vegetation, and ground.
        We have therefore decided to generate a new dataset of city maps from raw airborne \textit{light detection and ranging} (LiDAR) point clouds provided by the Geoportal Berlin \cite{geoportal_berlin}.
        Using the software LAStools \cite{lastools}, we have automatically separated the point clouds into the classes ground, building, and vegetation.
        The recognition of buildings has been further improved by incorporating building footprints from \cite{geoportal_berlin}, categorizing all elevated (above-ground) points inside the footprints as buildings.
        From there we extracted normalized digital surface models (nDSM), representing the height above the ground for the two classes buildings and vegetation. 
        Since Berlin shows a small deviation in elevation in most areas, we approximated the ground as a flat surface and normalized the heights of objects relative to it.

        \subsubsection{Radio Map Simulations}\label{sec:rm_simulations}
        The ground truth radio maps were generated with the GPU accelerated X3D propagation model in the widely used ray-tracing software Wireless InSite \cite{wi}.
        Since it was impossible to determine the exact materials of specific houses or types of trees, we have chosen standard material types for these upon import to the simulation software.
        The nDSMs have been converted to polygons for the simulations by grouping up neighboring pixels with approximately the same height and interpolating the boundaries.
        The ground and all buildings are assumed to represent solid structures, allowing reflections from surfaces and diffractions around edges but blocking transmissions.
        Vegetation on the other hand is modeled as a solely attenuating material.
        Exploiting that the exact shape of foliage objects is therefore less important, we have chosen less accurate interpolation for the vegetation layers upon constructing the polygons.
        By doing so, we could significantly reduce the runtime, which depends heavily on the number of faces in the environment geometry, and hence generate a larger dataset.

        To simulate a realistic cellular environment, Tx that model cellular base stations were placed on the edges of buildings at a height of $2$m from the roof and restricted to heights between $6$m and $30$m above the ground. 
        A dense grid of receivers with isotropic antennas and a spacing of $1$m at a height of $1.5$m was defined to model typical UE in the network, e.g. smartphones of people walking on the sidewalks or devices in cars.

        \begin{figure}[!htbp]
            \centering
            \begin{subfigure}[t]{0.1\textwidth}
            \includegraphics[width=\textwidth]{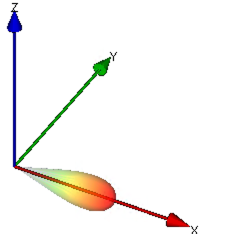}
            \caption{$(15^{\circ}, 30^{\circ})$}
            \label{fig:ant:narrow}
            \end{subfigure}
            \hfill
            \begin{subfigure}[t]{0.1\textwidth}
            \includegraphics[width=\textwidth]{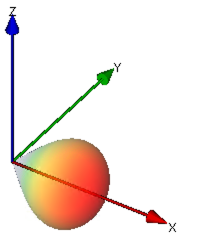}
            \caption{$(45^{\circ}, 90^{\circ})$}
            \label{fig:ant:middle}
            \end{subfigure}
            \hfill
            \begin{subfigure}[t]{0.1\textwidth}
            \includegraphics[width=\textwidth]{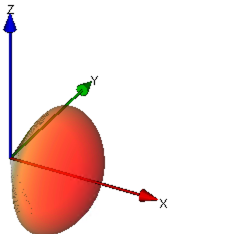}
            \caption{$(90^{\circ}, 120^{\circ})$}
            \label{fig:ant:wide}
            \end{subfigure}
            \caption{3D plot of antenna radiation patterns. The values in bracket indicate the half power beam width and first null beam width, respectively.}
            \label{fig:ant}
        \end{figure}

        For the Tx, we have selected the directional antenna type in the ray-tracing software.
        It allows to simulate an idealized main beam without any side lobes.
        By adjusting the parameters half power beam width and first null beam width, it can model very narrow beams or antennas covering wider sectors.
        For each Tx position, we test several combinations of azimuth and tilt angles in the direction pointing away from the building, on which the Tx is placed, with a line of sight algorithm.
        Orientations that do not cover at least a certain area on the ground are disregarded.
        For all angle combinations found, we select one of the narrow and one of the wide antenna patterns in Table \ref{table:antennas}.
        Although on each city map we consider different Tx positions and orientations, each of them is used in a separate simulation, resulting in a different radio map with a single Tx per data sample.

        \begin{table*}[!ht]
            \centering
            \begin{tabular}{|c|cccc|cccc|}
                \hline
                {Pattern}                        &   \multicolumn{4}{c|}{Narrow} &   \multicolumn{4}{c|}{Wide}   \\
                \hline
                Half power beam width   &   $15^{\circ}$  &   $15^{\circ}$  &   $30^{\circ}$ &   $45^{\circ}$  &   $15^{\circ}$  &   $30^{\circ}$  &   $45^{\circ}$  &   $90^{\circ}$  \\  
                \hline
                First null beam width   &   $30^{\circ}$  &   $60^{\circ}$  &   $60^{\circ}$  &   $60^{\circ}$  &   $90^{\circ}$  &   $90^{\circ}$  &   $90^{\circ}$  &   $120^{\circ}$ \\
                \hline
            \end{tabular}
            \caption{Antenna parameters.}
            \label{table:antennas}
        \end{table*}

        In Table \ref{table:simulation_parameters} we list further parameters.
        Note that only the values in the upper part of the table are used to configure the ray-tracing simulations, whereas the values below are calculated or assumed for post-processing, as described below.
        \begin{table}[!htbp]
            \centering
            \begin{tabular}{|c|c|}
                \hline
                Carrier frequency   &   $3.7$ GHz \\
                Number of reflections   &   2 \\
                Number of diffractions   &   1 \\
                Number of transmissions   &   0 \\
                \hline
                Maximum path loss & $-50$ dB \\
                Bandwidth  &   $10$ MHz \\
                Tx Input power  &   $23$ dBm \\
                Noise Power Spectral Density    &   $-174$dBm/Hz \\
                Noise figure    &   $0\dB$ \\
                \hline
            \end{tabular}
            \caption{Simulation and dataset parameters.}
            \label{table:simulation_parameters}
        \end{table}
        The simulation output are the magnitudes of the channel coefficients at each Rx on the grid, which we use to calculate the path loss \eqref{eq:power_ratio}.
        The path loss values are converted to $\dB$-scale and rescaled and cut following \cite{radiounet}.
        Assuming a bandwidth of $B=10\MHz$, thermal noise power spectral density $(N_0)_{\dBm/\Hz}=-174\dBm/\Hz$ and an idealistic noise figure of $(\NoiseFigure)_{\dB}=0\dB$ at the Rx (Table \ref{table:simulation_parameters}), the noise floor $\NoiseFloor$ is calculated as 
        \begin{equation*}
            (\NoiseFloor)_{\dB}  = 10\cdot\log_{10}(B) + (N_0)_{\dBm/\Hz} + (\NoiseFigure)_{\dB} = -127\dB.
        \end{equation*}
        We are interested in locations where the received signal satisfies a signal-to-noise ratio of at least $(\SNR)_{\dB} = 0\dB$, i.e. where the path loss is greater than or equal to the threshold
        \begin{equation*}
            (\PL_{\thr})_{\dB}   =   - (\power_{\Tx})_{\dBm} + (\SNR)_{\dB} + (\NoiseFloor)_{\dB} = -127\dB.
        \end{equation*}
        Any signal arriving with a path loss below this threshold is irrelevant in practice, but it may be beneficial for the model to also see values slightly below it during training in order to understand propagation phenomena better.
        Therefore, we cut off the path loss values at a second threshold $\PL_{\trnc}$, which is chosen so that 
        \begin{equation*}
            (\PL_{\max})_{\dB} - (\PL_{\thr})_{\dB} \approx 4\left((\PL_{\thr})_{\dB} - (\PL_{\trnc})_{\dB}\right),
        \end{equation*}
        where $(\PL_{\max})_{\dB}=-50\dB$ is the largest path loss value occurring across our dataset, which yields $(\PL_{\trnc})_{\dB}=-147\dB$.
        Finally, the values are mapped with an affine-linear transformation to the interval $[0,1]$ so that $\PL_{\trnc}$ and $\PL_{\max}$ correspond to $0$ and $1$, respectively, to obtain the radio map as a grayscale image.
        The rescaling assures that, in contrast do $\dB$-scale, the most relevant parts of the radio map containing strong signal dominate parts with very low signal in terms of magnitude.

        \subsubsection{Other Data}\label{sec:other_data}
        We also include images from \cite{geoportal_bb} taken about $2$ months after the LiDAR measurements, cut and downscaled to match the position and resolution of the nDSMs.
        Furthermore, the dataset contains 2D polygons with height attributes extracted from the nDSMs, which have been used for the ray-tracing simulations.
        Files containing line-of-sight information for each radio map are also included but not considered in this work.
    
    \subsection{Experiment Design}\label{sec:experiments}
        \subsubsection{CNN-Architectures}\label{sec:architectures}
            As a lightweight baseline model, we use the RadioUNet \cite{radiounet}, more concretely the first part of the WNet described by the authors.
            In principle, it follows the structure of the original UNet \cite{radiounet}, but it features more down and upsampling layers and, in some parts, convolutions with a larger kernel size, allowing to propagate information over longer distances.
            As a second baseline, we include experiments with the PMNet proposed in \cite{pmnet}, featuring a relatively deep encoder consisting of stacked ResNet-layers and several parallel convolutional layers with varying {dilation} \cite{dilated} after the encoder.
            PMNet has been shown to perform better than RadioUNet and other architectures on different datasets \cite{pmnet}, \cite{challenge}.
            These were the only architectures designed for the radio map prediction task available in the literature, for which the code has been made publicly available.

            \tikzset{every picture/.style={line width=0.75pt}} 
            \begin{figure}[!htbp]
                \centering
                \resizebox{0.8\linewidth}{!}{
                \begin{tikzpicture}[x=0.75pt,y=0.75pt,yscale=-1,xscale=1]

                \draw  [draw opacity=0] (111,98) -- (212.5,98) -- (212.5,199) -- (111,199) -- cycle ; \draw   (111,98) -- (111,199)(131,98) -- (131,199)(151,98) -- (151,199)(171,98) -- (171,199)(191,98) -- (191,199)(211,98) -- (211,199) ; \draw   (111,98) -- (212.5,98)(111,118) -- (212.5,118)(111,138) -- (212.5,138)(111,158) -- (212.5,158)(111,178) -- (212.5,178)(111,198) -- (212.5,198) ; \draw    ;
                \draw  [draw opacity=0] (244,97) -- (345.5,97) -- (345.5,198) -- (244,198) -- cycle ; \draw   (244,97) -- (244,198)(264,97) -- (264,198)(284,97) -- (284,198)(304,97) -- (304,198)(324,97) -- (324,198)(344,97) -- (344,198) ; \draw   (244,97) -- (345.5,97)(244,117) -- (345.5,117)(244,137) -- (345.5,137)(244,157) -- (345.5,157)(244,177) -- (345.5,177)(244,197) -- (345.5,197) ; \draw    ;
                \draw  [draw opacity=0] (371,97) -- (472.5,97) -- (472.5,198) -- (371,198) -- cycle ; \draw   (371,97) -- (371,198)(391,97) -- (391,198)(411,97) -- (411,198)(431,97) -- (431,198)(451,97) -- (451,198)(471,97) -- (471,198) ; \draw   (371,97) -- (472.5,97)(371,117) -- (472.5,117)(371,137) -- (472.5,137)(371,157) -- (472.5,157)(371,177) -- (472.5,177)(371,197) -- (472.5,197) ; \draw    ;
                \draw  [fill={rgb, 255:red, 208; green, 2; blue, 27 }  ,fill opacity=1 ] (131,118) -- (151,118) -- (151,138) -- (131,138) -- cycle ;
                \draw  [fill={rgb, 255:red, 208; green, 2; blue, 27 }  ,fill opacity=1 ] (244,137) -- (264,137) -- (264,157) -- (244,157) -- cycle ;
                \draw  [fill={rgb, 255:red, 208; green, 2; blue, 27 }  ,fill opacity=1 ] (284,97) -- (304,97) -- (304,117) -- (284,117) -- cycle ;
                \draw  [fill={rgb, 255:red, 208; green, 2; blue, 27 }  ,fill opacity=1 ] (324,137) -- (344,137) -- (344,157) -- (324,157) -- cycle ;
                \draw  [fill={rgb, 255:red, 208; green, 2; blue, 27 }  ,fill opacity=1 ] (284,137) -- (304,137) -- (304,157) -- (284,157) -- cycle ;
                \draw  [fill={rgb, 255:red, 208; green, 2; blue, 27 }  ,fill opacity=1 ] (171,158) -- (191,158) -- (191,178) -- (171,178) -- cycle ;
                \draw  [fill={rgb, 255:red, 208; green, 2; blue, 27 }  ,fill opacity=1 ] (171,138) -- (191,138) -- (191,158) -- (171,158) -- cycle ;
                \draw  [fill={rgb, 255:red, 208; green, 2; blue, 27 }  ,fill opacity=1 ] (151,158) -- (171,158) -- (171,178) -- (151,178) -- cycle ;
                \draw  [fill={rgb, 255:red, 208; green, 2; blue, 27 }  ,fill opacity=1 ] (131,158) -- (151,158) -- (151,178) -- (131,178) -- cycle ;
                \draw  [fill={rgb, 255:red, 208; green, 2; blue, 27 }  ,fill opacity=1 ] (131,138) -- (151,138) -- (151,158) -- (131,158) -- cycle ;
                \draw  [fill={rgb, 255:red, 208; green, 2; blue, 27 }  ,fill opacity=1 ] (171,118) -- (191,118) -- (191,138) -- (171,138) -- cycle ;
                \draw  [fill={rgb, 255:red, 208; green, 2; blue, 27 }  ,fill opacity=1 ] (151,118) -- (171,118) -- (171,138) -- (151,138) -- cycle ;
                \draw  [fill={rgb, 255:red, 208; green, 2; blue, 27 }  ,fill opacity=1 ] (151,138) -- (171,138) -- (171,158) -- (151,158) -- cycle ;
                \draw  [fill={rgb, 255:red, 208; green, 2; blue, 27 }  ,fill opacity=1 ] (284,177) -- (304,177) -- (304,197) -- (284,197) -- cycle ;
                \draw  [fill={rgb, 255:red, 208; green, 2; blue, 27 }  ,fill opacity=1 ] (244,97) -- (264,97) -- (264,117) -- (244,117) -- cycle ;
                \draw  [fill={rgb, 255:red, 208; green, 2; blue, 27 }  ,fill opacity=1 ] (244,177) -- (264,177) -- (264,197) -- (244,197) -- cycle ;
                \draw  [fill={rgb, 255:red, 208; green, 2; blue, 27 }  ,fill opacity=1 ] (324,177) -- (344,177) -- (344,197) -- (324,197) -- cycle ;
                \draw  [fill={rgb, 255:red, 208; green, 2; blue, 27 }  ,fill opacity=1 ] (324,97) -- (344,97) -- (344,117) -- (324,117) -- cycle ;
                \draw  [fill={rgb, 255:red, 208; green, 2; blue, 27 }  ,fill opacity=1 ] (371,177) -- (391,177) -- (391,197) -- (371,197) -- cycle ;
                \draw  [fill={rgb, 255:red, 208; green, 2; blue, 27 }  ,fill opacity=1 ] (371,137) -- (391,137) -- (391,157) -- (371,157) -- cycle ;
                \draw  [fill={rgb, 255:red, 208; green, 2; blue, 27 }  ,fill opacity=1 ] (411,177) -- (431,177) -- (431,197) -- (411,197) -- cycle ;
                \draw  [fill={rgb, 255:red, 208; green, 2; blue, 27 }  ,fill opacity=1 ] (431,177) -- (451,177) -- (451,197) -- (431,197) -- cycle ;
                \draw  [fill={rgb, 255:red, 208; green, 2; blue, 27 }  ,fill opacity=1 ] (451,137) -- (471,137) -- (471,157) -- (451,157) -- cycle ;
                \draw  [fill={rgb, 255:red, 208; green, 2; blue, 27 }  ,fill opacity=1 ] (431,117) -- (451,117) -- (451,137) -- (431,137) -- cycle ;
                \draw  [fill={rgb, 255:red, 208; green, 2; blue, 27 }  ,fill opacity=1 ] (391,97) -- (411,97) -- (411,117) -- (391,117) -- cycle ;
                \draw  [fill={rgb, 255:red, 208; green, 2; blue, 27 }  ,fill opacity=1 ] (371,97) -- (391,97) -- (391,117) -- (371,117) -- cycle ;
                \draw  [fill={rgb, 255:red, 208; green, 2; blue, 27 }  ,fill opacity=1 ] (451,117.5) -- (471,117.5) -- (471,137.5) -- (451,137.5) -- cycle ;
                \draw    (275,127.5) -- (255.43,108.4) ;
                \draw [shift={(254,107)}, rotate = 44.31] [color={rgb, 255:red, 0; green, 0; blue, 0 }  ][line width=0.75]    (10.93,-3.29) .. controls (6.95,-1.4) and (3.31,-0.3) .. (0,0) .. controls (3.31,0.3) and (6.95,1.4) .. (10.93,3.29)   ;
                \draw    (294,168.5) -- (294,185) ;
                \draw [shift={(294,187)}, rotate = 270] [color={rgb, 255:red, 0; green, 0; blue, 0 }  ][line width=0.75]    (10.93,-3.29) .. controls (6.95,-1.4) and (3.31,-0.3) .. (0,0) .. controls (3.31,0.3) and (6.95,1.4) .. (10.93,3.29)   ;
                \draw    (315.5,147.75) -- (332,147.08) ;
                \draw [shift={(334,147)}, rotate = 177.68] [color={rgb, 255:red, 0; green, 0; blue, 0 }  ][line width=0.75]    (10.93,-3.29) .. controls (6.95,-1.4) and (3.31,-0.3) .. (0,0) .. controls (3.31,0.3) and (6.95,1.4) .. (10.93,3.29)   ;
                \draw    (316,168) -- (332.62,185.55) ;
                \draw [shift={(334,187)}, rotate = 226.55] [color={rgb, 255:red, 0; green, 0; blue, 0 }  ][line width=0.75]    (10.93,-3.29) .. controls (6.95,-1.4) and (3.31,-0.3) .. (0,0) .. controls (3.31,0.3) and (6.95,1.4) .. (10.93,3.29)   ;
                \draw    (275,167.5) -- (255.47,185.64) ;
                \draw [shift={(254,187)}, rotate = 317.12] [color={rgb, 255:red, 0; green, 0; blue, 0 }  ][line width=0.75]    (10.93,-3.29) .. controls (6.95,-1.4) and (3.31,-0.3) .. (0,0) .. controls (3.31,0.3) and (6.95,1.4) .. (10.93,3.29)   ;
                \draw    (315,127.5) -- (332.64,108.47) ;
                \draw [shift={(334,107)}, rotate = 132.83] [color={rgb, 255:red, 0; green, 0; blue, 0 }  ][line width=0.75]    (10.93,-3.29) .. controls (6.95,-1.4) and (3.31,-0.3) .. (0,0) .. controls (3.31,0.3) and (6.95,1.4) .. (10.93,3.29)   ;
                \draw    (274.5,147.25) -- (256,147.02) ;
                \draw [shift={(254,147)}, rotate = 0.7] [color={rgb, 255:red, 0; green, 0; blue, 0 }  ][line width=0.75]    (10.93,-3.29) .. controls (6.95,-1.4) and (3.31,-0.3) .. (0,0) .. controls (3.31,0.3) and (6.95,1.4) .. (10.93,3.29)   ;
                \draw    (294.5,127.25) -- (294.05,109) ;
                \draw [shift={(294,107)}, rotate = 88.59] [color={rgb, 255:red, 0; green, 0; blue, 0 }  ][line width=0.75]    (10.93,-3.29) .. controls (6.95,-1.4) and (3.31,-0.3) .. (0,0) .. controls (3.31,0.3) and (6.95,1.4) .. (10.93,3.29)   ;
                \draw    (441,127) -- (459,127.45) ;
                \draw [shift={(461,127.5)}, rotate = 181.43] [color={rgb, 255:red, 0; green, 0; blue, 0 }  ][line width=0.75]    (10.93,-3.29) .. controls (6.95,-1.4) and (3.31,-0.3) .. (0,0) .. controls (3.31,0.3) and (6.95,1.4) .. (10.93,3.29)   ;
                \draw    (402,167) -- (382.45,185.62) ;
                \draw [shift={(381,187)}, rotate = 316.4] [color={rgb, 255:red, 0; green, 0; blue, 0 }  ][line width=0.75]    (10.93,-3.29) .. controls (6.95,-1.4) and (3.31,-0.3) .. (0,0) .. controls (3.31,0.3) and (6.95,1.4) .. (10.93,3.29)   ;
                \draw    (421,147) -- (383,147) ;
                \draw [shift={(381,147)}, rotate = 360] [color={rgb, 255:red, 0; green, 0; blue, 0 }  ][line width=0.75]    (10.93,-3.29) .. controls (6.95,-1.4) and (3.31,-0.3) .. (0,0) .. controls (3.31,0.3) and (6.95,1.4) .. (10.93,3.29)   ;
                \draw    (401,147) -- (381.89,108.79) ;
                \draw [shift={(381,107)}, rotate = 63.43] [color={rgb, 255:red, 0; green, 0; blue, 0 }  ][line width=0.75]    (10.93,-3.29) .. controls (6.95,-1.4) and (3.31,-0.3) .. (0,0) .. controls (3.31,0.3) and (6.95,1.4) .. (10.93,3.29)   ;
                \draw    (421,167) -- (421,185) ;
                \draw [shift={(421,187)}, rotate = 270] [color={rgb, 255:red, 0; green, 0; blue, 0 }  ][line width=0.75]    (10.93,-3.29) .. controls (6.95,-1.4) and (3.31,-0.3) .. (0,0) .. controls (3.31,0.3) and (6.95,1.4) .. (10.93,3.29)   ;
                \draw    (440,167) -- (440.9,185) ;
                \draw [shift={(441,187)}, rotate = 267.14] [color={rgb, 255:red, 0; green, 0; blue, 0 }  ][line width=0.75]    (10.93,-3.29) .. controls (6.95,-1.4) and (3.31,-0.3) .. (0,0) .. controls (3.31,0.3) and (6.95,1.4) .. (10.93,3.29)   ;
                \draw    (441,146) -- (459,146.9) ;
                \draw [shift={(461,147)}, rotate = 182.86] [color={rgb, 255:red, 0; green, 0; blue, 0 }  ][line width=0.75]    (10.93,-3.29) .. controls (6.95,-1.4) and (3.31,-0.3) .. (0,0) .. controls (3.31,0.3) and (6.95,1.4) .. (10.93,3.29)   ;
                \draw    (401,127) -- (401,109) ;
                \draw [shift={(401,107)}, rotate = 90] [color={rgb, 255:red, 0; green, 0; blue, 0 }  ][line width=0.75]    (10.93,-3.29) .. controls (6.95,-1.4) and (3.31,-0.3) .. (0,0) .. controls (3.31,0.3) and (6.95,1.4) .. (10.93,3.29)   ;
                \draw    (421,127) -- (439,127) ;
                \draw [shift={(441,127)}, rotate = 180] [color={rgb, 255:red, 0; green, 0; blue, 0 }  ][line width=0.75]    (10.93,-3.29) .. controls (6.95,-1.4) and (3.31,-0.3) .. (0,0) .. controls (3.31,0.3) and (6.95,1.4) .. (10.93,3.29)   ;
                \end{tikzpicture}
                }
                \caption{Illustration (inspired by \cite{dcn}) of sampling points in standard, dilated and deformable convolution with kernel size $3\times3$.}
                \label{fig:convs}
            \end{figure}
            
\begin{figure}
                \tikzset{every picture/.style={line width=0.75pt}} 
                \resizebox{\linewidth}{!}{

\tikzset{every picture/.style={line width=0.75pt}} 

\begin{tikzpicture}[x=0.75pt,y=0.75pt,yscale=-1,xscale=1]

\draw  [fill={rgb, 255:red, 155; green, 155; blue, 155 }  ,fill opacity=1 ] (35,90) -- (40,90) -- (40,145) -- (35,145) -- cycle ;
\draw  [fill={rgb, 255:red, 74; green, 144; blue, 226 }  ,fill opacity=1 ] (40,116.25) -- (51.75,116.25) -- (51.75,115) -- (60,117.5) -- (51.75,120) -- (51.75,118.75) -- (40,118.75) -- cycle ;
\draw  [color={rgb, 255:red, 0; green, 0; blue, 0 }  ,draw opacity=1 ][fill={rgb, 255:red, 65; green, 117; blue, 5 }  ,fill opacity=1 ] (65,116.25) -- (77,116.25) -- (77,115) -- (85,117.5) -- (77,120) -- (77,118.75) -- (65,118.75) -- cycle ;
\draw  [color={rgb, 255:red, 0; green, 0; blue, 0 }  ,draw opacity=1 ][fill={rgb, 255:red, 208; green, 2; blue, 27 }  ,fill opacity=1 ] (88.75,145) -- (88.75,156.42) -- (90,156.42) -- (87.5,164.04) -- (85,156.42) -- (86.25,156.42) -- (86.25,145) -- cycle ;
\draw  [color={rgb, 255:red, 0; green, 0; blue, 0 }  ,draw opacity=1 ][fill={rgb, 255:red, 208; green, 2; blue, 27 }  ,fill opacity=1 ] (113.75,220) -- (113.75,232) -- (115,232) -- (112.5,240) -- (110,232) -- (111.25,232) -- (111.25,220) -- cycle ;
\draw   (125,254.5) .. controls (125,254.22) and (125.22,254) .. (125.5,254) .. controls (125.78,254) and (126,254.22) .. (126,254.5) .. controls (126,254.78) and (125.78,255) .. (125.5,255) .. controls (125.22,255) and (125,254.78) .. (125,254.5) -- cycle ;
\draw   (134,264.5) .. controls (134,264.78) and (134.22,265) .. (134.5,265) .. controls (134.78,265) and (135,264.78) .. (135,264.5) .. controls (135,264.22) and (134.78,264) .. (134.5,264) .. controls (134.22,264) and (134,264.22) .. (134,264.5) -- cycle ;
\draw   (144,274.5) .. controls (144,274.22) and (144.22,274) .. (144.5,274) .. controls (144.78,274) and (145,274.22) .. (145,274.5) .. controls (145,274.78) and (144.78,275) .. (144.5,275) .. controls (144.22,275) and (144,274.78) .. (144,274.5) -- cycle ;
\draw    (80,85) -- (320,85) ;
\draw  [color={rgb, 255:red, 0; green, 0; blue, 0 }  ,draw opacity=1 ][fill={rgb, 255:red, 65; green, 117; blue, 5 }  ,fill opacity=1 ] (90,191.25) -- (100.43,191.25) -- (100.43,190) -- (110,192.5) -- (100.43,195) -- (100.43,193.75) -- (90,193.75) -- cycle ;
\draw  [fill={rgb, 255:red, 74; green, 144; blue, 226 }  ,fill opacity=1 ] (5,415.25) -- (17,415.25) -- (17,414) -- (25,416.5) -- (17,419) -- (17,417.75) -- (5,417.75) -- cycle ;
\draw  [color={rgb, 255:red, 0; green, 0; blue, 0 }  ,draw opacity=1 ][fill={rgb, 255:red, 65; green, 117; blue, 5 }  ,fill opacity=1 ] (5,440.25) -- (16.5,440.25) -- (16.5,439) -- (25,441.5) -- (16.5,444) -- (16.5,442.75) -- (5,442.75) -- cycle ;
\draw  [color={rgb, 255:red, 0; green, 0; blue, 0 }  ,draw opacity=1 ][fill={rgb, 255:red, 208; green, 2; blue, 27 }  ,fill opacity=1 ] (284.75,425) -- (284.75,437) -- (286,437) -- (283.5,445) -- (281,437) -- (282.25,437) -- (282.25,425) -- cycle ;
\draw  [color={rgb, 255:red, 0; green, 0; blue, 0 }  ,draw opacity=1 ][fill={rgb, 255:red, 189; green, 16; blue, 224 }  ,fill opacity=1 ] (282.25,470) -- (282.25,458) -- (281,458) -- (283.5,450) -- (286,458) -- (284.75,458) -- (284.75,470) -- cycle ;
\draw    (105,160) -- (295,160) ;
\draw    (271,412) -- (289,411.1) ;
\draw [shift={(291,411)}, rotate = 177.14] [color={rgb, 255:red, 0; green, 0; blue, 0 }  ][line width=0.75]    (10.93,-3.29) .. controls (6.95,-1.4) and (3.31,-0.3) .. (0,0) .. controls (3.31,0.3) and (6.95,1.4) .. (10.93,3.29)   ;
\draw  [color={rgb, 255:red, 0; green, 0; blue, 0 }  ,draw opacity=1 ][fill={rgb, 255:red, 245; green, 166; blue, 35 }  ,fill opacity=1 ] (5,464.5) -- (17,464.5) -- (17,463) -- (25,466) -- (17,469) -- (17,467.5) -- (5,467.5) -- cycle ;
\draw  [fill={rgb, 255:red, 155; green, 155; blue, 155 }  ,fill opacity=1 ] (60,90) -- (65,90) -- (65,145) -- (60,145) -- cycle ;
\draw  [fill={rgb, 255:red, 155; green, 155; blue, 155 }  ,fill opacity=1 ] (85,90) -- (90,90) -- (90,145) -- (85,145) -- cycle ;
\draw  [fill={rgb, 255:red, 155; green, 155; blue, 155 }  ,fill opacity=1 ] (85,165) -- (90,165) -- (90,220) -- (85,220) -- cycle ;
\draw  [fill={rgb, 255:red, 155; green, 155; blue, 155 }  ,fill opacity=1 ] (110,165) -- (115,165) -- (115,220) -- (110,220) -- cycle ;
\draw  [fill={rgb, 255:red, 155; green, 155; blue, 155 }  ,fill opacity=1 ] (160,260) -- (165,260) -- (165,315) -- (160,315) -- cycle ;
\draw  [color={rgb, 255:red, 0; green, 0; blue, 0 }  ,draw opacity=1 ][fill={rgb, 255:red, 208; green, 2; blue, 27 }  ,fill opacity=1 ] (188.75,315) -- (188.75,327) -- (190,327) -- (187.5,335) -- (185,327) -- (186.25,327) -- (186.25,315) -- cycle ;
\draw  [color={rgb, 255:red, 0; green, 0; blue, 0 }  ,draw opacity=1 ][fill={rgb, 255:red, 65; green, 117; blue, 5 }  ,fill opacity=1 ] (165,286.25) -- (175.43,286.25) -- (175.43,285) -- (185,287.5) -- (175.43,290) -- (175.43,288.75) -- (165,288.75) -- cycle ;
\draw  [fill={rgb, 255:red, 155; green, 155; blue, 155 }  ,fill opacity=1 ] (185,260) -- (190,260) -- (190,315) -- (185,315) -- cycle ;
\draw  [fill={rgb, 255:red, 155; green, 155; blue, 155 }  ,fill opacity=1 ] (185,335) -- (190,335) -- (190,390) -- (185,390) -- cycle ;
\draw  [color={rgb, 255:red, 0; green, 0; blue, 0 }  ,draw opacity=1 ][fill={rgb, 255:red, 245; green, 166; blue, 35 }  ,fill opacity=1 ] (190,361.25) -- (202,361.25) -- (202,360) -- (210,362.5) -- (202,365) -- (202,363.75) -- (190,363.75) -- cycle ;
\draw  [fill={rgb, 255:red, 155; green, 155; blue, 155 }  ,fill opacity=1 ] (210,335) -- (215,335) -- (215,390) -- (210,390) -- cycle ;
\draw  [color={rgb, 255:red, 0; green, 0; blue, 0 }  ,draw opacity=1 ][fill={rgb, 255:red, 245; green, 166; blue, 35 }  ,fill opacity=1 ] (220,361.25) -- (232,361.25) -- (232,360) -- (240,362.5) -- (232,365) -- (232,363.75) -- (220,363.75) -- cycle ;
\draw  [fill={rgb, 255:red, 155; green, 155; blue, 155 }  ,fill opacity=1 ] (215,335) -- (220,335) -- (220,390) -- (215,390) -- cycle ;
\draw  [fill={rgb, 255:red, 155; green, 155; blue, 155 }  ,fill opacity=1 ] (240,335) -- (245,335) -- (245,390) -- (240,390) -- cycle ;
\draw  [color={rgb, 255:red, 0; green, 0; blue, 0 }  ,draw opacity=1 ][fill={rgb, 255:red, 189; green, 16; blue, 224 }  ,fill opacity=1 ] (241.25,335) -- (241.25,323) -- (240,323) -- (242.5,315) -- (245,323) -- (243.75,323) -- (243.75,335) -- cycle ;
\draw  [fill={rgb, 255:red, 155; green, 155; blue, 155 }  ,fill opacity=1 ] (240,260) -- (245,260) -- (245,315) -- (240,315) -- cycle ;
\draw  [color={rgb, 255:red, 0; green, 0; blue, 0 }  ,draw opacity=1 ][fill={rgb, 255:red, 245; green, 166; blue, 35 }  ,fill opacity=1 ] (245,286.25) -- (257,286.25) -- (257,285) -- (265,287.5) -- (257,290) -- (257,288.75) -- (245,288.75) -- cycle ;
\draw  [fill={rgb, 255:red, 155; green, 155; blue, 155 }  ,fill opacity=1 ] (235,260) -- (240,260) -- (240,315) -- (235,315) -- cycle ;
\draw   (310.18,255) .. controls (310.46,255) and (310.68,255.22) .. (310.69,255.49) .. controls (310.69,255.77) and (310.47,256) .. (310.19,256) .. controls (309.92,256) and (309.69,255.78) .. (309.69,255.51) .. controls (309.68,255.23) and (309.9,255) .. (310.18,255) -- cycle ;
\draw   (300.28,264.15) .. controls (300,264.16) and (299.78,264.39) .. (299.79,264.66) .. controls (299.79,264.94) and (300.02,265.16) .. (300.3,265.15) .. controls (300.57,265.15) and (300.79,264.92) .. (300.79,264.65) .. controls (300.78,264.37) and (300.56,264.15) .. (300.28,264.15) -- cycle ;
\draw   (290.49,275.31) .. controls (290.77,275.3) and (291,275.52) .. (291,275.8) .. controls (291,276.08) and (290.78,276.3) .. (290.51,276.31) .. controls (290.23,276.31) and (290,276.09) .. (290,275.82) .. controls (290,275.54) and (290.22,275.31) .. (290.49,275.31) -- cycle ;
\draw  [fill={rgb, 255:red, 155; green, 155; blue, 155 }  ,fill opacity=1 ] (340,165) -- (345,165) -- (345,220) -- (340,220) -- cycle ;
\draw  [color={rgb, 255:red, 0; green, 0; blue, 0 }  ,draw opacity=1 ][fill={rgb, 255:red, 189; green, 16; blue, 224 }  ,fill opacity=1 ] (341.25,165) -- (341.25,153) -- (340,153) -- (342.5,145) -- (345,153) -- (343.75,153) -- (343.75,165) -- cycle ;
\draw  [fill={rgb, 255:red, 155; green, 155; blue, 155 }  ,fill opacity=1 ] (340,90) -- (345,90) -- (345,145) -- (340,145) -- cycle ;
\draw  [color={rgb, 255:red, 0; green, 0; blue, 0 }  ,draw opacity=1 ][fill={rgb, 255:red, 74; green, 144; blue, 226 }  ,fill opacity=1 ] (345,116.25) -- (357,116.25) -- (357,115) -- (365,117.5) -- (357,120) -- (357,118.75) -- (345,118.75) -- cycle ;
\draw  [fill={rgb, 255:red, 155; green, 155; blue, 155 }  ,fill opacity=1 ] (335,90) -- (340,90) -- (340,145) -- (335,145) -- cycle ;
\draw  [fill={rgb, 255:red, 155; green, 155; blue, 155 }  ,fill opacity=1 ] (365,90) -- (370,90) -- (370,145) -- (365,145) -- cycle ;
\draw  [fill={rgb, 255:red, 155; green, 155; blue, 155 }  ,fill opacity=1 ] (315,165) -- (320,165) -- (320,220) -- (315,220) -- cycle ;
\draw  [color={rgb, 255:red, 0; green, 0; blue, 0 }  ,draw opacity=1 ][fill={rgb, 255:red, 245; green, 166; blue, 35 }  ,fill opacity=1 ] (320,191.25) -- (332,191.25) -- (332,190) -- (340,192.5) -- (332,195) -- (332,193.75) -- (320,193.75) -- cycle ;
\draw  [fill={rgb, 255:red, 155; green, 155; blue, 155 }  ,fill opacity=1 ] (310,165) -- (315,165) -- (315,220) -- (310,220) -- cycle ;
\draw    (105,160) -- (105,185) ;
\draw    (90,185) -- (105,185) ;
\draw    (295,160) -- (295,185) ;
\draw    (295,185) -- (308,185) ;
\draw [shift={(310,185)}, rotate = 180] [color={rgb, 255:red, 0; green, 0; blue, 0 }  ][line width=0.75]    (10.93,-3.29) .. controls (6.95,-1.4) and (3.31,-0.3) .. (0,0) .. controls (3.31,0.3) and (6.95,1.4) .. (10.93,3.29)   ;
\draw    (80,85) -- (80,110) ;
\draw    (65,110) -- (80,110) ;
\draw    (205,330) -- (205,355) ;
\draw    (190,355) -- (205,355) ;
\draw    (320,85) -- (320,110) ;
\draw    (320,110) -- (333,110) ;
\draw [shift={(335,110)}, rotate = 180] [color={rgb, 255:red, 0; green, 0; blue, 0 }  ][line width=0.75]    (10.93,-3.29) .. controls (6.95,-1.4) and (3.31,-0.3) .. (0,0) .. controls (3.31,0.3) and (6.95,1.4) .. (10.93,3.29)   ;
\draw    (215,330) -- (205,330) ;
\draw    (215,330) -- (215,333) ;
\draw [shift={(215,335)}, rotate = 270] [color={rgb, 255:red, 0; green, 0; blue, 0 }  ][line width=0.75]    (10.93,-3.29) .. controls (6.95,-1.4) and (3.31,-0.3) .. (0,0) .. controls (3.31,0.3) and (6.95,1.4) .. (10.93,3.29)   ;
\draw    (180,255) -- (220,255) ;
\draw    (180,255) -- (180,280) ;
\draw    (165,280) -- (180,280) ;
\draw    (220,255) -- (220,280) ;
\draw    (220,280) -- (233,280) ;
\draw [shift={(235,280)}, rotate = 180] [color={rgb, 255:red, 0; green, 0; blue, 0 }  ][line width=0.75]    (10.93,-3.29) .. controls (6.95,-1.4) and (3.31,-0.3) .. (0,0) .. controls (3.31,0.3) and (6.95,1.4) .. (10.93,3.29)   ;
\draw  [fill={rgb, 255:red, 155; green, 155; blue, 155 }  ,fill opacity=1 ] (262.5,260) -- (267.5,260) -- (267.5,315) -- (262.5,315) -- cycle ;
\draw  [color={rgb, 255:red, 0; green, 0; blue, 0 }  ,draw opacity=1 ][fill={rgb, 255:red, 189; green, 16; blue, 224 }  ,fill opacity=1 ] (316.25,240) -- (316.25,228) -- (315,228) -- (317.5,220) -- (320,228) -- (318.75,228) -- (318.75,240) -- cycle ;

\draw (77,67.4) node [anchor=north west][inner sep=0.75pt]  [font=\footnotesize]  {$2C$};
\draw (102,67.4) node [anchor=north west][inner sep=0.75pt]  [font=\footnotesize]  {$4C$};
\draw (2.4,388) node [anchor=north west][inner sep=0.75pt]  [font=\footnotesize,rotate=-270]  {$\frac{H}{2^{d}} \times \frac{W}{2^{d}}$};
\draw (2.4,216) node [anchor=north west][inner sep=0.75pt]  [font=\footnotesize,rotate=-270]  {$\frac{H}{2} \times \frac{W}{2}$};
\draw (7.4,140) node [anchor=north west][inner sep=0.75pt]  [font=\footnotesize,rotate=-270]  {$H\times W$};
\draw (186,67.4) node [anchor=north west][inner sep=0.75pt]  [font=\footnotesize]  {$2^{d} C$};
\draw (31,409) node [anchor=north west][inner sep=0.75pt]  [font=\footnotesize] [align=left] {2x(Conv, BN, ReLU, res)};
\draw (32,435) node [anchor=north west][inner sep=0.75pt]  [font=\footnotesize] [align=left] {Conv, Bn, ReLU, DC, BN, ReLU, res};
\draw (299,430) node [anchor=north west][inner sep=0.75pt]  [font=\footnotesize] [align=left] {MaxPool};
\draw (297,455) node [anchor=north west][inner sep=0.75pt]  [font=\footnotesize] [align=left] {ConvT, ReLU};
\draw (26,67.4) node [anchor=north west][inner sep=0.75pt]  [font=\footnotesize]  {$C_{in}$};
\draw (56,67.4) node [anchor=north west][inner sep=0.75pt]  [font=\footnotesize]  {$C$};
\draw (146,67.4) node [anchor=north west][inner sep=0.75pt]  [font=\footnotesize]  {$2^{d-1} C$};
\draw (297,407) node [anchor=north west][inner sep=0.75pt]  [font=\footnotesize] [align=left] {Copy, cat};
\draw (33,460) node [anchor=north west][inner sep=0.75pt]  [font=\footnotesize] [align=left] {2x(Conv, BN, ReLU, res), DC, BN, ReLU, res};
\draw (2.4,324) node [anchor=north west][inner sep=0.75pt]  [font=\footnotesize,rotate=-270]  {$\frac{H}{2^{d-1}} \times \frac{W}{2^{d-1}}$};
\draw (221,67.4) node [anchor=north west][inner sep=0.75pt]  [font=\footnotesize]  {$2^{d-1} C$};
\draw (333,67.4) node [anchor=north west][inner sep=0.75pt]  [font=\footnotesize]  {$C$};
\draw (355,67.4) node [anchor=north west][inner sep=0.75pt]  [font=\footnotesize]  {$C_{out}$};
\draw (306,67.4) node [anchor=north west][inner sep=0.75pt]  [font=\footnotesize]  {$2C$};
\node at (340, 10) {\includegraphics[width=0.2\columnwidth]{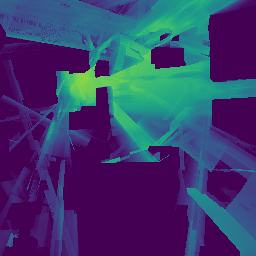}};
\node at (70, 0) {\includegraphics[width=0.2\columnwidth]{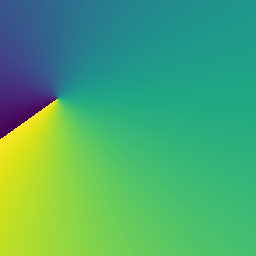}};
\node at (90, 10) {\includegraphics[width=0.2\columnwidth]{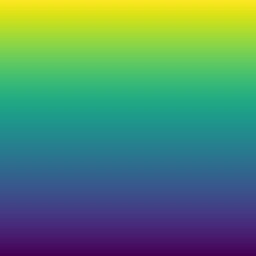}};
\node at (110, 20) {\includegraphics[width=0.2\columnwidth]{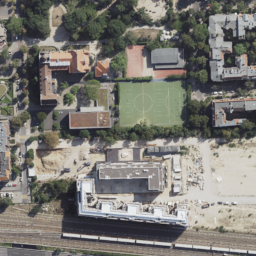}};

\end{tikzpicture}

            }
            \caption{Outline of the UNetDCN architecture. 
            The numbers on the left describe the spatial dimensions and the numbers on top the channel dimension of the tensors, represented as gray blocks. 
            In our experiments, $H=W=256$, $C_{out}=1$, $C_{in}$ depends on the chosen inputs and the width and depth hyperparameters are set to $C=32, d=3$.
            Blocks of network layers are depicted by arrows, and explained below.
            The channel dimension is always changed by the first convolution inside of each block.
            The layers used are standard 3$\times$3 convolutions (\textit{Conv}), batch normalization (\textit{BN}), rectified linear unit (\textit{ReLU}), residual connections (\textit{res}), concatenation of tensors along the channel dimension (\textit{cat}), downsampling with $2\times2$ MaxPool and upsampling with $2\times2$ transposed convolutions (ConvT).}
            \label{fig:dcn_arch}
            \end{figure}

            We propose the use of a model dubbed UNetDCN that combines the structure of UNet \cite{unet} with deformable convolutions.
            Originally presented in \cite{dcn}, this CNN layer has been used in several computer vision problems, but, to the best of our knowledge, we are the first ones to apply it for radio map prediction.
            Similar to a dilated convolution, it allows to enlarge the receptive field by sampling the input at positions further away. 
            The sampling points are not fixed  (Fig. \ref{fig:convs}), instead, the offset compared to a standard convolution is computed from the input with learnable parameters.
            Intuitively, this should make it easier to propagate information in arbitrary directions compared to dilated convolutions.
            Furthermore, in comparison to standard convolutions with a large kernel size, deformable convolutions require less parameters and numeric operations.
            The architecture is illustrated in Fig. \ref{fig:dcn_arch} and the exact implementation can be found on our GitHub page mentioned in Section \ref{sec:introduction}.
            Many aspects of the network are similar to the original UNet \cite{unet}, such as repeated convolutions increasing the channel dimension and max pooling reducing the spatial dimension in the encoder part,
            upsampling and simultaneous reduction of the channel dimension in the decoder part and skip connections from the encoder to the decoder at several stages to preserve information.
            The main changes on the other hand are the replacement of some standard convolutions by deformable convolutions, additional batch normalization and residual connections.
            In Section \ref{sec:ablation}, we provide an ablation study investigating the choice of the hyperparameters in Fig. \ref{fig:dcn_arch} and the use of deformable convolutions.

            Table \ref{table:complexity} compares the complexity of the different architectures across various aspects. 
            PMNet exhibits the highest complexity in terms of multiply-accumulate operations (MACs) and number of parameters.
            The proposed UNetDCN has the smallest number of parameters, resulting in a minimal memory footprint, while its MAC count is slightly higher than that of RadioUNet.

            \begin{table}
                \centering
                \begin{tabular}{|c|c|c|c|}
                    \hline
                    Model                   &   RadioUNet \cite{radiounet}    &   PMNet   \cite{pmnet}      &   UNetDCN    \\
                    \hline
                    \hline
                    \#Params$^{a}$     &  10.9M                         &   33.4M                     &  4.5M \\
                    \#MACs$^{a}$           &  7.7G                           &   50.7G                     &  10.3G \\
                    \hline
                \end{tabular}
                \caption{Complexity of the considered architectures for the default inputs and batch size 1.$^{a}$Calculated using DeepSpeed \cite{deepspeed}.}
                \label{table:complexity}
            \end{table}

        \subsubsection{Input Features}\label{sec:input_features}
\begin{figure*}[!htbp]
                \centering
                \begin{subfigure}[t]{0.1\linewidth}
                    \includegraphics[width=\linewidth]{jaens9}
                    \caption{}
                    \label{fig:input_features:target}
                \end{subfigure}
                \begin{subfigure}[t]{0.1\linewidth}
                    \includegraphics[width=\linewidth]{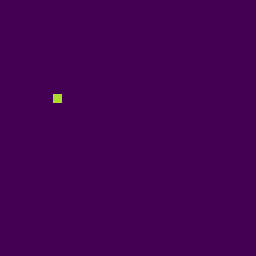}
                    \caption{}
                    \label{fig:input_features:tx_one_hot}
                \end{subfigure}
                \begin{subfigure}[t]{0.1\linewidth}
                    \includegraphics[width=\linewidth]{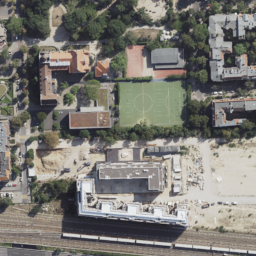}
                    \caption{}
                    \label{fig:input_features:img}
                \end{subfigure}
                \begin{subfigure}[t]{0.1\linewidth}
                    \includegraphics[width=\linewidth]{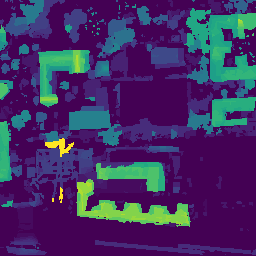}
                    \caption{}
                    \label{fig:input_features:ndsm_all}
                \end{subfigure}
                \begin{subfigure}[t]{0.1\linewidth}
                    \includegraphics[width=\linewidth]{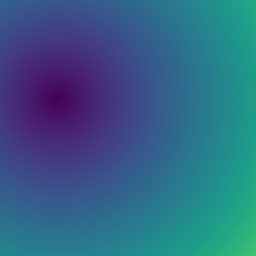}
                    \caption{}
                    \label{fig:input_features:dist2d}
                \end{subfigure}
                \begin{subfigure}[t]{0.1\linewidth}
                    \includegraphics[width=\linewidth]{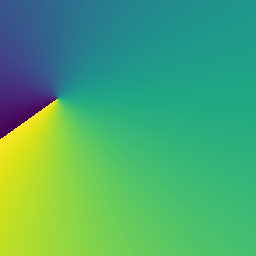}
                    \caption{}
                    \label{fig:input_features:azimuth}
                \end{subfigure}
                \begin{subfigure}[t]{0.1\linewidth}
                    \includegraphics[width=\linewidth]{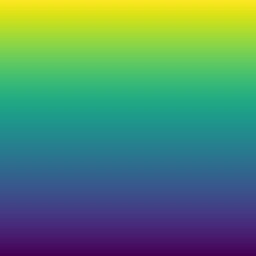}
                    \caption{}
                    \label{fig:input_features:gax}
                \end{subfigure}
                \begin{subfigure}[t]{0.1\linewidth}
                    \includegraphics[width=\linewidth]{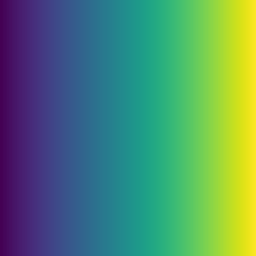}
                    \caption{}
                    \label{fig:input_features:gay}
                \end{subfigure}
                \begin{subfigure}[t]{0.1\linewidth}
                    \includegraphics[width=\linewidth]{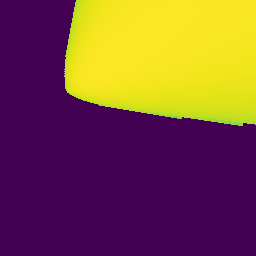}
                    \caption{}
                    \label{fig:input_features:gain_floor}
                \end{subfigure}
                \caption{Example radio map \subref{fig:input_features:target} and (normalized) input features. 
                    \subref{fig:input_features:tx_one_hot} Tx location (visibility enhanced, in reality only marked by one pixel), 
                    \subref{fig:input_features:img} aerial image, 
                    \subref{fig:input_features:ndsm_all} nDSM, 
                    \subref{fig:input_features:dist2d} 2D distance to Tx, 
                    \subref{fig:input_features:azimuth} azimuth angle,
                    \subref{fig:input_features:gax}, 
                    \subref{fig:input_features:gay} map coordinates from GA \cite{radiotrans},
                    \subref{fig:input_features:gain_floor} Tx antenna pattern projected onto the ground.}
                \label{fig:input_features}
            \end{figure*}

            Following the related literature (Section \ref{sec:related_works}), we aim to encode all relevant parameters that change between the different simulations in 2D images, with each pixel representing a specific location on the map.
            In the following, we provide an explanation of the input features we consider.
            Several examples for a sample from the dataset are shown in Fig. \ref{fig:input_features}, and further details of the implementation can be found in the code.
            All inputs are normalized to values in $[-1, 1]$ before being fed to the CNN.

            In the $2D$ setting, the positions of the Tx, buildings, and potentially other objects are often given in a \textit{one-hot encoding} in separate binary tensors, where a $1$ indicates the presence of the Tx or object and a $0$ absence in the location corresponding to each pixel (see e.g. \cite{radiounet}).
            Since in our work the height of the Tx, buildings, and vegetation are relevant, we assign these as the values to the pixels instead, as it is done in \cite{radiotrans} or \cite{fadenet} for example (Fig. \ref{fig:input_features}\subref{fig:input_features:tx_one_hot}-\subref{fig:input_features:gay}).
            The sparsity of the tensor representing the Tx location is potentially problematic, as the standard layers used in CNNs inherently have a very limited field of view, and it therefore takes several convolutional and downsampling operations to spread the information to other parts of the map.

            The authors of \cite{radiotrans} propose to tackle this issue by making the information about the spatial position of each pixel together with the location of the Tx explicitly available to the model in the form of constant input tensors for the $x, y$ and $z$ coordinates of the Tx and two tensors showing the $x$ and $y$ coordinates of each pixel.
            Although they state that their main intention was to find an alternative to the usual positional embedding in vision transformer layers, we have found that this approach also improves the performance of other CNN models.
            They denote their idea as {Grid Anchor (GA)}.
            Additionally, we provide the azimuth angle between the direction the Tx is pointing at and the straight line to each point of the map, as in \cite{plnet}.
            This encodes Tx directivity and helps determine whether two objects lie along the same path.
            Lastly, we also include the distance in the $x-y$ plane from the Tx to each point as in \cite{qiu22}.
            All these inputs related to positions on the maps are together denoted as \textit{coords} in Sec. \ref{sec:numerical_results}.

            To link the antenna pattern to the spatial positions, we use spherical coordinates centered in the Tx location and rotated according to the Tx orientation to look up the gain in dB corresponding to azimuth and tilt angle for each point on the ground, as it is done in \cite{plnet} as well.

            As described in the introduction, precise information about the locations, heights, and shapes of buildings, and especially vegetation is scarce.
            Aerial imagery, on the other hand, is an, in many cases, easier accessible data source which at least partially contains this information implicitly.
            We perform experiments to see to what extent the models are capable of predicting the radio map from just images or potentially with additional sight information.
            Intuitively, this requires the models to implicitly perform a semantic segmentation, i.e. pixel-wise classification, of the input image in order to find objects relevant for the signal propagation and to estimate the heights of the found objects.
            The images contain, besides the usual RGB channels, an infrared channel, which we include by default. 
            Performance without it is also tested (\textit{w/o IR}).
            Lastly, we additionally provide the networks with an unclassified nDSM depicting elevation of buildings and vegetation together, as this kind of information is easier to acquire than height maps for each class individually.

    \section{RESULTS ON RADIO MAP PREDICTION}\label{sec:numerical_results}

        \begin{table*}
            \centering
            \begin{tabular}{|c||c|c|c|c|c|c|}
                \hline
                Model                  &   \multicolumn{2}{c|}{RadioUNet \cite{radiounet}}  &   \multicolumn{2}{c|}{PMNet \cite{pmnet}}    &   \multicolumn{2}{c|}{UNetDCN}   \\
                \hline
                \hline
                Input                   &   RMSE                &   NMSE                    &   RMSE                    &   NMSE                &   RMSE                    &   NMSE  \\
                \hline
                Image                   &   $\textit{0.094} (0.073/0.093)$   &   $\textit{0.0038}$      &   $\textit{0.091} (0.068/0.089)$       &  $\textit{0.0036}$   & \textit{\textbf{0.088}} (0.077/0.086) & \textit{\textbf{0.0033}} \\
                Image w/o IR            &   $0.097 (0.070/0.094)$            &   $0.0040$               &   $0.093 (0.078/0.092)$                &  $0.0037$           & \textbf{0.089} (0.074/0.085) & \textbf{0.0034} \\ 
                Image + coords          &   $0.098 (0.076/0.096)$            &   $0.0041$               &   $0.092 (0.070/0.090)$                &  $0.0036$             &  \textbf{0.089} (0.072/0.085) & \textbf{0.0034} \\
                \hline
                Image + nDSM            &   $0.076 (0.062/0.071)$            &   $0.0025$               &   ${0.070} (0.057/0.066)$       &  ${0.0021}$   & \textbf{0.069} (0.061/0.065) & \textbf{0.0020}  \\
                Image + nDSM + coords   &   $\underline{0.073} (0.063/0.070)$&   $\underline{0.0023}$   &   ${\underline{0.068}} (0.060/0.065)$&  ${\underline{0.0020}}$&  \underline{\textbf{0.067}} (0.061/0.064) & \underline{\textbf{0.0019}} \\
                \hline
            \end{tabular}
            \caption{Errors on the test set (in brackets on the training/validation set). Lowest test errors in each column (input) without nDSM in \textit{italics}, with nDSM \underline{underlined} and per row (model) in \textbf{bold}.}
            \label{table:img}
        \end{table*}

        We use about $80\%$ of the samples as the training set, $10\%$ for validation, and $10\%$ for testing, ensuring that the city maps do not overlap between the different sets.
        The validation set is applied to first reduce the learning rate and later stop training when the loss stagnates, and to determine the best model weights to save.
        The test set is only used at the very end to evaluate the final performance and generalization capability.
        During training, we apply random flips and rotations as data augmentations.

        All models are trained with respect to mean-squared error (MSE) between the predicted and ground truth radio maps.
        We measure the final performance in terms of root-mean-square error (RMSE) in grayscale.
        Recall that the grayscale radio map values in $[0,1]$ stem from an affine transformation of the possible path loss values in dB, which lie in $[-127\text{dB}, -50\text{dB}]$, hence, the RMSE in dB can be obtained from multiplying the RMSE in grayscale by a factor of $77$.
        Furthermore, we report the normalized mean-squared error (NMSE) after conversion to dB scale, which emphasizes the importance of samples with a high signal power (see \cite{radiounet}).

        Our code is implemented in PyTorch Lightning. 
        All models are trained with a batch size of $32$ using the Adam optimizer with an initial learning rate of $10^{-4}$ on A100 GPUs.
        Typically, the training is stopped due to stagnating validation loss within the first $40$ epochs.
        The reported losses are generated on the test set.

        \begin{figure}
            \includegraphics{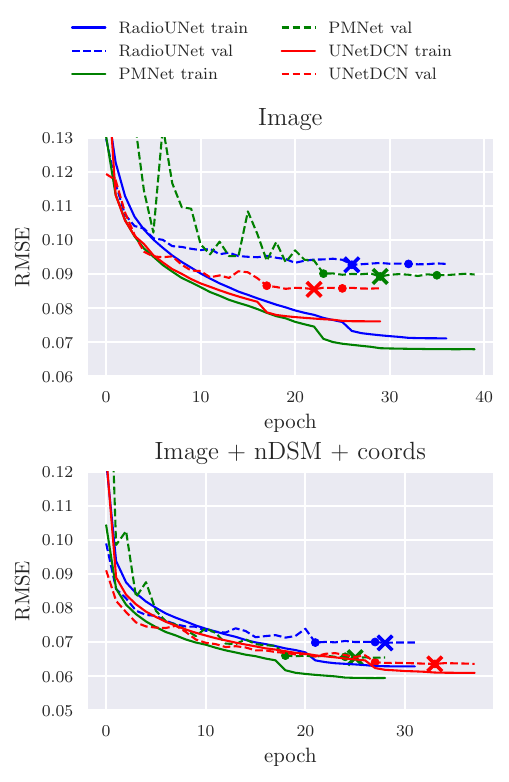}
            \caption{Errors per epoch on the training and on the validation set. Dots mark epochs in which the learning rate is reduced due to stagnating validation loss. The final model weights stem from the epoch with the lowest validation loss, which is marked with a cross.}
            \label{fig:loss_curves}
        \end{figure}
        \begin{figure*}[!htbp]
            \centering
            \begin{subfigure}[t]{0.1\textwidth}
            \includegraphics[width=\textwidth]{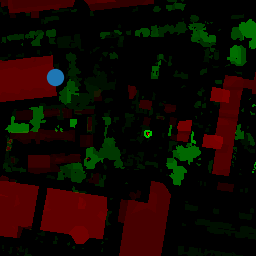}
            \end{subfigure}
            \begin{subfigure}[t]{0.1\textwidth}
            \includegraphics[width=\textwidth]{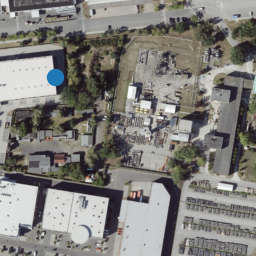}
            \end{subfigure}
            \begin{subfigure}[t]{0.1\textwidth}
                \includegraphics[width=\textwidth]{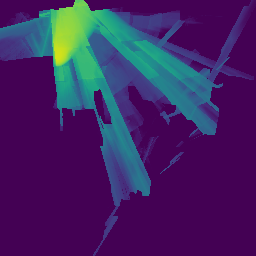}
            \end{subfigure}
            \begin{subfigure}[t]{0.1\textwidth}
            \includegraphics[width=\textwidth]{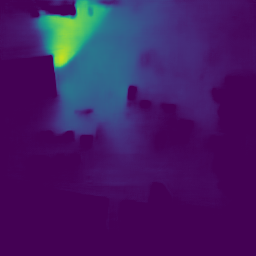}
            \caption*{0.018}
            \end{subfigure}
            \begin{subfigure}[t]{0.1\textwidth}
            \includegraphics[width=\textwidth]{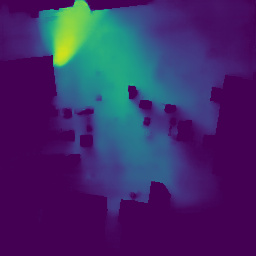}
            \caption*{0.015}
            \end{subfigure}
            \begin{subfigure}[t]{0.1\textwidth}
            \includegraphics[width=\textwidth]{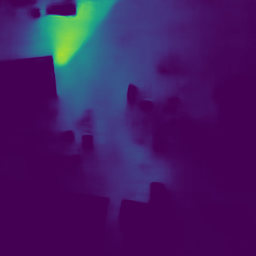}
            \caption*{0.019}
            \end{subfigure}
            \begin{subfigure}[t]{0.1\textwidth}
            \includegraphics[width=\textwidth]{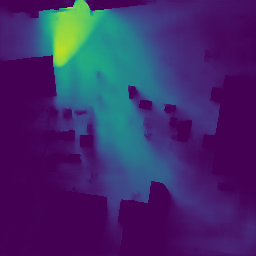}
            \caption*{0.013}
            \end{subfigure}
            \begin{subfigure}[t]{0.1\textwidth}
            \includegraphics[width=\textwidth]{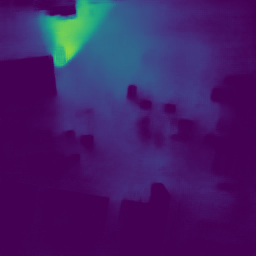}
            \caption*{0.018}
            \end{subfigure}
            \begin{subfigure}[t]{0.1\textwidth}
            \includegraphics[width=\textwidth]{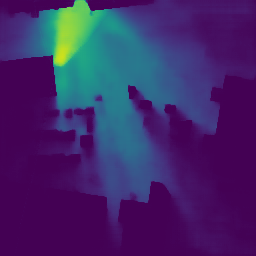}
            \caption*{0.009}
            \end{subfigure}
            \begin{subfigure}[t]{0.028\textwidth}
                \includegraphics[width=\textwidth]{jaens5}
            \end{subfigure}
            \\
            \centering
            \begin{subfigure}[t]{0.1\textwidth}
            \includegraphics[width=\textwidth]{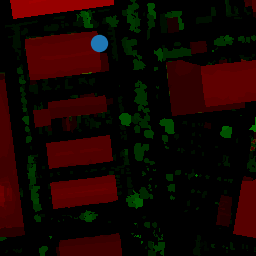}
            \end{subfigure}
            \begin{subfigure}[t]{0.1\textwidth}
            \includegraphics[width=\textwidth]{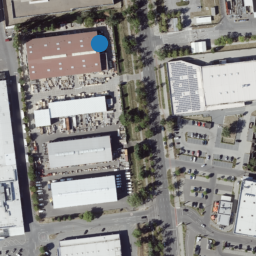}
            \end{subfigure}
            \begin{subfigure}[t]{0.1\textwidth}
                \includegraphics[width=\textwidth]{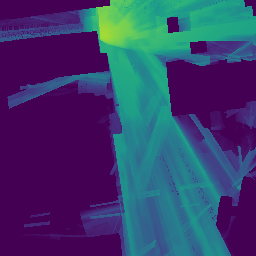}
            \end{subfigure}
            \begin{subfigure}[t]{0.1\textwidth}
            \includegraphics[width=\textwidth]{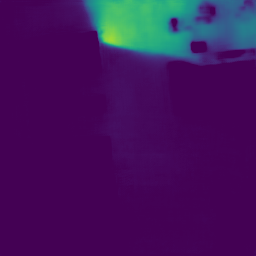}
            \caption*{0.056}
            \end{subfigure}
            \begin{subfigure}[t]{0.1\textwidth}
            \includegraphics[width=\textwidth]{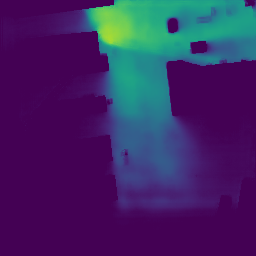}
            \caption*{0.024}
            \end{subfigure}
            \begin{subfigure}[t]{0.1\textwidth}
            \includegraphics[width=\textwidth]{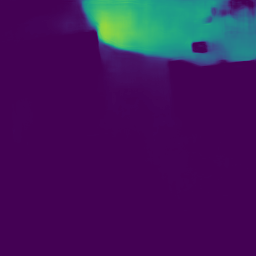}
            \caption*{0.056}
            \end{subfigure}
            \begin{subfigure}[t]{0.1\textwidth}
            \includegraphics[width=\textwidth]{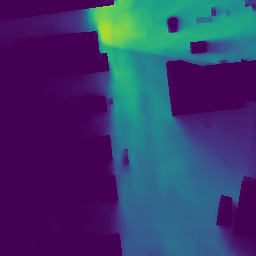}
            \caption*{0.006}
            \end{subfigure}
            \begin{subfigure}[t]{0.1\textwidth}
            \includegraphics[width=\textwidth]{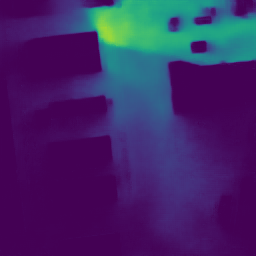}
            \caption*{0.029}
            \end{subfigure}
            \begin{subfigure}[t]{0.1\textwidth}
            \includegraphics[width=\textwidth]{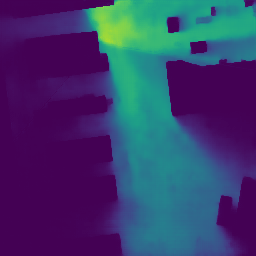}
            \caption*{0.006}
            \end{subfigure}
            \begin{subfigure}[t]{0.028\textwidth}
                \includegraphics[width=\textwidth]{jaens5}
            \end{subfigure}
            \\
            \centering
            \begin{subfigure}[t]{0.1\textwidth}
            \includegraphics[width=\textwidth]{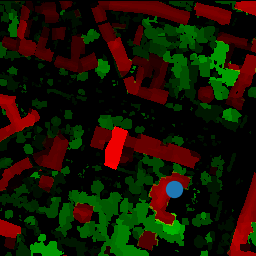}
            \end{subfigure}
            \begin{subfigure}[t]{0.1\textwidth}
            \includegraphics[width=\textwidth]{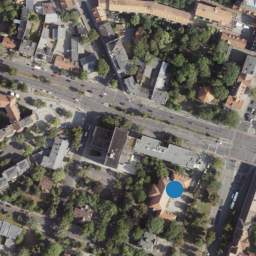}
            \end{subfigure}
            \begin{subfigure}[t]{0.1\textwidth}
                \includegraphics[width=\textwidth]{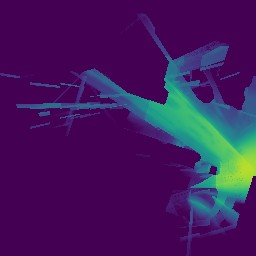}
            \end{subfigure}
            \begin{subfigure}[t]{0.1\textwidth}
            \includegraphics[width=\textwidth]{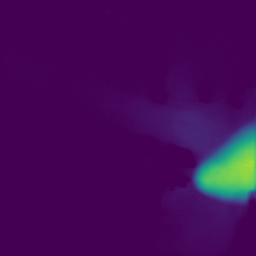}
            \caption*{0.023}
            \end{subfigure}
            \begin{subfigure}[t]{0.1\textwidth}
            \includegraphics[width=\textwidth]{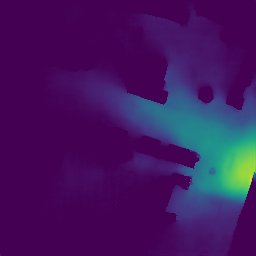}
            \caption*{0.007}
            \end{subfigure}
            \begin{subfigure}[t]{0.1\textwidth}
            \includegraphics[width=\textwidth]{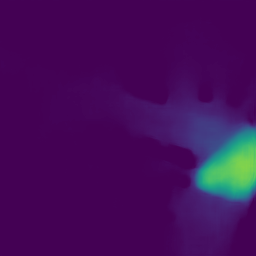}
            \caption*{0.020}
            \end{subfigure}
            \begin{subfigure}[t]{0.1\textwidth}
            \includegraphics[width=\textwidth]{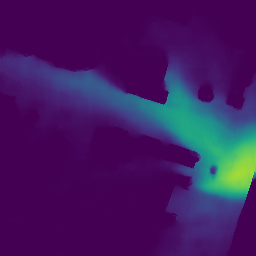}
            \caption*{0.005}
            \end{subfigure}
            \begin{subfigure}[t]{0.1\textwidth}
            \includegraphics[width=\textwidth]{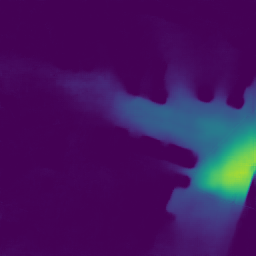}
            \caption*{0.010}
            \end{subfigure}
            \begin{subfigure}[t]{0.1\textwidth}
            \includegraphics[width=\textwidth]{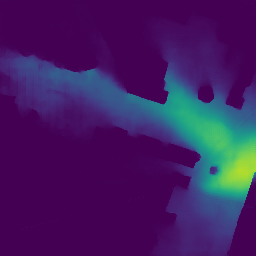}
            \caption*{0.006}
            \end{subfigure}
            \begin{subfigure}[t]{0.028\textwidth}
                \includegraphics[width=\textwidth]{jaens5}
            \end{subfigure}
            \\
            \centering
            \begin{subfigure}[t]{0.1\textwidth}
            \includegraphics[width=\textwidth]{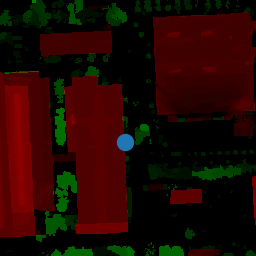}
            \caption{}
            \label{fig:comparison_models_img:gis}
            \end{subfigure}
            \begin{subfigure}[t]{0.1\textwidth}
            \includegraphics[width=\textwidth]{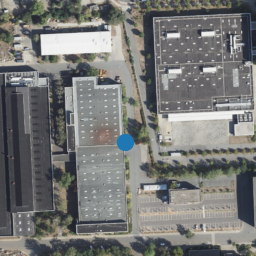}
            \caption{}
            \label{fig:comparison_models_img:img}
            \end{subfigure}
            \begin{subfigure}[t]{0.1\textwidth}
                \includegraphics[width=\textwidth]{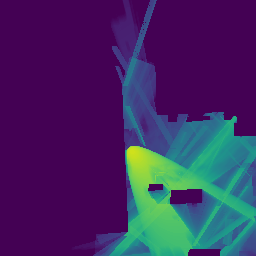}
                \caption{}
                \label{fig:comparison_models_img:target}
            \end{subfigure}
            \begin{subfigure}[t]{0.1\textwidth}
            \includegraphics[width=\textwidth]{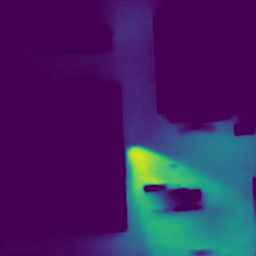}
            \caption{0.009}
            \label{fig:comparison_models_img:radiounet}
            \end{subfigure}
            \begin{subfigure}[t]{0.1\textwidth}
            \includegraphics[width=\textwidth]{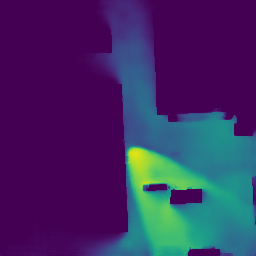}
            \caption*{0.005}
            \end{subfigure}
            \begin{subfigure}[t]{0.1\textwidth}
            \includegraphics[width=\textwidth]{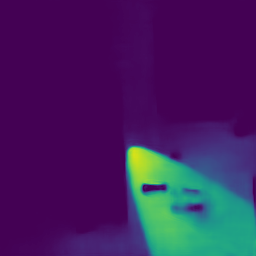}
            \caption{0.019}
            \label{fig:comparison_models_img:pm}
            \end{subfigure}
            \begin{subfigure}[t]{0.1\textwidth}
            \includegraphics[width=\textwidth]{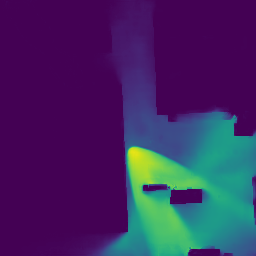}
            \caption*{0.004}
            \end{subfigure}
            \begin{subfigure}[t]{0.1\textwidth}
            \includegraphics[width=\textwidth]{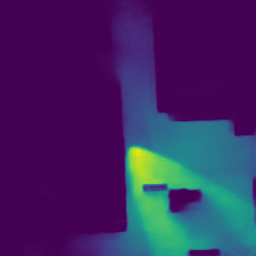}
            \caption{0.006}
            \label{fig:comparison_models_img:dcn}
            \end{subfigure}
            \begin{subfigure}[t]{0.1\textwidth}
            \includegraphics[width=\textwidth]{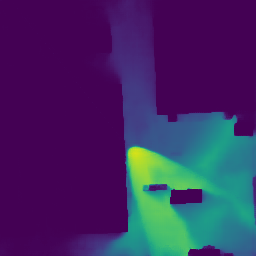}
            \caption*{0.003}
            \end{subfigure}
            \begin{subfigure}[t]{0.028\textwidth}
                \includegraphics[width=\textwidth]{jaens5}
            \end{subfigure}
            \caption{Examples of predictions --
            \subref{fig:comparison_models_img:gis} overlayed height maps with buildings in red, vegetation in green and Tx position in blue,
            \subref{fig:comparison_models_img:img} aerial image,
            \subref{fig:comparison_models_img:target} ground truth radio map, predictions only from images and from images, nDSMs and coords by: 
                \subref{fig:comparison_models_img:radiounet} RadioUNet \cite{radiounet},
                \subref{fig:comparison_models_img:pm} PMNet \cite{pmnet},
                \subref{fig:comparison_models_img:dcn} UNetDCN, MSE below.}
            \label{fig:comparison_models_img}
        \end{figure*}

        In Table \ref{table:img}, we list the results.
        Overall, all models are able to estimate the radio maps with very good accuracy between $5.2$ and $7.2$ dB RMSE.
        Removing the infrared channel from the input image causes a slight decrease in accuracy.
        The height map improves the predictions significantly for all models, as expected, since inferring the height of objects from a two-dimensional image is inherently ill-posed \cite{im2height}. 
        Across all models, the additional inputs related to the positions on the map provide a slight improvement when also the height map is used.
        However, they deteriorate performance when only the image is provided, which is rather surprising.
        A possible explanation could be that the height information, map coordinates, and distances share the same physical units, while the channels of the aerial images contain a completely different type of information.
        We observe that, for all inputs considered, the more lightweight RadioUNet performs slightly worse than the two other models.
        The UNetDCN we propose achieves comparable accuracy when height information is included and outperforms the baselines when height data is absent.

        The numeric errors appear relatively close.
        Visual inspection of the predictions reveals that our dataset contains diverse sample types. 
        For some samples, there are clear differences between models and/or inputs, while other predictions appear very similar both visually and in terms of loss.
        In Fig. \ref{fig:comparison_models_img}, we provide a visual comparison of the predictions of the models with and without access to height information for a few exemplary samples.
        While all models generate satisfactory predictions in most cases and correctly recognize buildings, trees, and the encoded antenna orientation and pattern, PMNet and UNetDCN appear to have an advantage over the RadioUNet when it comes to predicting long reflections, as in the first and third rows in Fig. \ref{fig:comparison_models_img}.
        We attribute this to the dilated and deformable convolutional layers, increasing the receptive field size.

\begin{figure*}[!htbp]
            \centering
            \begin{subfigure}[t]{0.1\textwidth}
                \includegraphics[width=\textwidth]{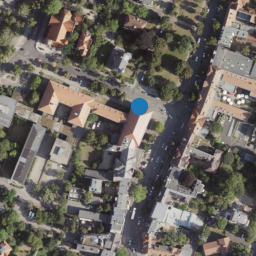}
            \end{subfigure}
            \begin{subfigure}[t]{0.1\textwidth}
                \includegraphics[width=\textwidth]{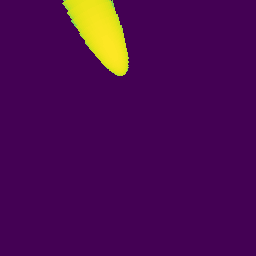}
            \end{subfigure}
            \begin{subfigure}[t]{0.028\textwidth}
                \includegraphics[width=\textwidth]{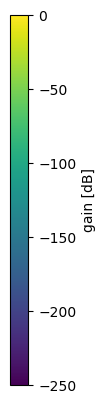}
            \end{subfigure}
            \begin{subfigure}[t]{0.1\textwidth}
            \includegraphics[width=\textwidth]{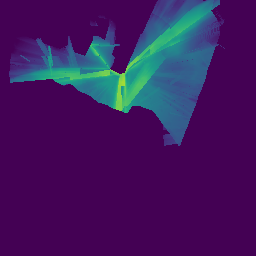}
            \end{subfigure}
            \begin{subfigure}[t]{0.1\textwidth}
            \includegraphics[width=\textwidth]{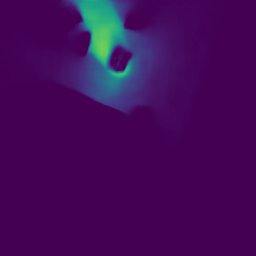}
            \caption*{0.033}
            \end{subfigure}
            \begin{subfigure}[t]{0.1\textwidth}
            \includegraphics[width=\textwidth]{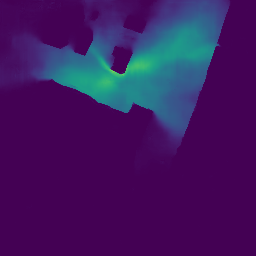}
            \caption*{0.005}
            \end{subfigure}
            \begin{subfigure}[t]{0.028\textwidth}
                \includegraphics[width=\textwidth]{jaens5}
            \end{subfigure}
            \\
            \begin{subfigure}[t]{0.1\textwidth}
                \includegraphics[width=\textwidth]{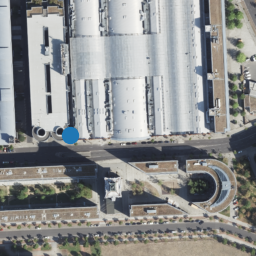}
                \caption{}
                \label{fig:comparison_images:img}
            \end{subfigure}
            \begin{subfigure}[t]{0.1\textwidth}
            \includegraphics[width=\textwidth]{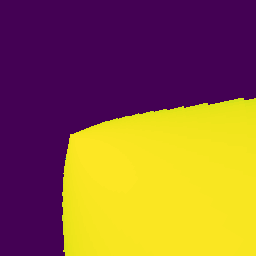}
            \caption{}
            \label{fig:comparison_images:gain}
            \end{subfigure}
            \begin{subfigure}[t]{0.028\textwidth}
                \includegraphics[width=\textwidth]{jaens95}
            \end{subfigure}
            \begin{subfigure}[t]{0.1\textwidth}
            \includegraphics[width=\textwidth]{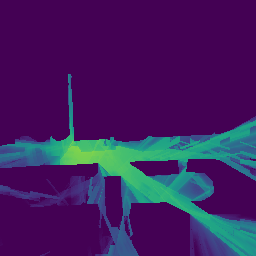}
            \caption{}
            \label{fig:comparison_images:target}
            \end{subfigure}
            \begin{subfigure}[t]{0.1\textwidth}
            \includegraphics[width=\textwidth]{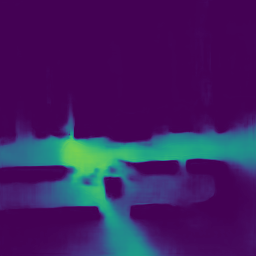}
            \caption{0.027}
            \label{fig:comparison_images:predimg}
            \end{subfigure}
            \begin{subfigure}[t]{0.1\textwidth}
            \includegraphics[width=\textwidth]{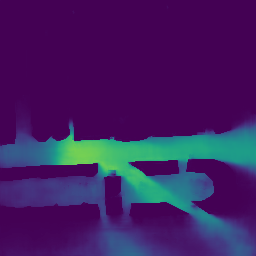}
            \caption{0.006}
            \label{fig:comparison_images:predimgndsm}
            \end{subfigure}
            \begin{subfigure}[t]{0.028\textwidth}
                \includegraphics[width=\textwidth]{jaens5}
            \end{subfigure}
            \caption{Examples of prediction, where height and positions of buildings are not determined correctly without nDSM. 
            \subref{fig:comparison_images:img} aerial image and Tx position in blue 
            \subref{fig:comparison_images:gain} antenna gain projected onto the floor 
            \subref{fig:comparison_images:target} ground truth radio map, 
            predictions by PMNet \cite{pmnet} with inputs: 
                \subref{fig:comparison_images:predimg} only image 
                \subref{fig:comparison_images:predimgndsm} image and unclassified nDSM, MSE below.}
            \label{fig:comparison_images_wrong}
        \end{figure*}

        In Fig. \ref{fig:loss_curves}, we plot the averaged training and validation losses for each epoch in the two most relevant scenarios.
        We observe that without height information, the gap between training and validation loss increases stronger, which implies overfitting.
        The reason for this could be that inferring the correct height of objects and their precise locations from just an image is very hard and the images do not contain sufficient information for stronger generalization.
        UNetDCN demonstrates better resistance to overfitting, as shown by the smaller gap between training and validation/test loss.
        In the first row of Fig. \ref{fig:comparison_images_wrong}, for example, we can see that the network without access to height information does not recognize that the house in the main lobe completely blocks the signal.
        Below in the second row, it seems to misinterpret the exact shape of the buildings due to shadows and predicts a non-existent propagation path.
        
    \subsection{Ablation Study}\label{sec:ablation}
        We have performed additional experiments to investigate the effect of the complexity of the proposed UNetDCN in Fig. \ref{fig:dcn_arch} on the achieved accuracy and to evaluate the usage of deformable convolutions (DC).
        For the two most relevant input features, Table \ref{table:ablation} lists the training and validation error together with the complexity in terms of the number of parameters and MACs for the baseline model and modified versions.
        
        Reducing the model complexity, either by decreasing depth ($d$) or width (number of channels, $C$), increases the validation loss notably by up to $0.004$.
        On the other hand, increasing the number of channels ($C$) only slightly improves validation loss at the cost of a significant increase in computational demand.
        The gap between training and validation loss is much larger when height information is absent. 
        This could potentially mean a higher risk of overfitting.
        For deeper networks ($d = 3,4$), we observe slight improvements when predicting from images alone. 
        However, when height information is included, validation loss remains the same or slightly worsens.
        Overall, the proposed parameters $d=3$ and $C=32$ provide a good validation accuracy at a low complexity.
        Increasing the size of the model could lead to marginal improvements but at the cost of significantly higher computational complexity and, in some cases, potentially a larger risk of overfitting.

        To assess the impact of deformable convolutions (DC), we also experiment with models following the same structure illustrated in Fig. \ref{fig:dcn_arch}, but with all DC layers replaced by other modules.
        We consider two alternatives: 
        First, standard convolutions with larger kernel sizes and second, the atrous spatial pyramid pooling (ASPP) block from \cite{pmnet}. 
        The ASPP block is an inception module with parallel $3\times3$ convolutions at different dilation rates (Fig. \ref{fig:convs}).
        With standard convolutions, even with large kernel sizes it is not possible to reach the same accuracy as with DC, despite a high complexity.
        Increasing the kernel size beyond $9\times9$ provides no further benefit but rather leads to performance deterioration.
        Also the ASPP block does not work well as a drop-in replacement in our proposed architecture.
        As we observe signs of potential overfitting in the first column of Table \ref{table:ablation}, we also try two more variants with reduced complexity in terms of the parameters $d$ and $C$, but the performance is still subpar compared to DC.
        These findings support our hypothesis that DC are a very efficient way to increase the receptive field in the radio map prediction task.

        \begin{table}\centering
            \begin{tabular}{|c||c|c|c|c|}
                \hline
                    &   \multicolumn{2}{c|}{RMSE} & & \\
                \cline{2-3}
                \multirow{1}{*}{Modification}   & \multirow{2}{*}{Image} & +nDSM & \multirow{1}{*}{\#Params}    & \multirow{1}{*}{\#MACs} \\
                 & & +coords & & \\
                \hline\hline
                -  & 0.077/0.086 & 0.061/0.064 & 4.5M & 10.3G \\
                $C=16$  & 0.082/0.090 & 0.066/0.067 & 1.19 M  & 3.15G \\
                $C=64$  & 0.069/0.085 & 0.059/0.063 & 17.56M & 36.9G \\
                $d=2$  & 0.081/0.088 & 0.067/0.068 & 1.15M & 8.02G \\
                $d=4$  & 0.073/0.085  & 0.062/0.064  & 17.7M & 12.6G \\
                $d=5$  & 0.072/0.084 & 0.063/0.065 & 69.8M & 14.7G \\
                \hline\hline
                Conv $5\times5$  & 0.083/0.090 & 0.067/0.070 & 7.05M & 31.3G \\
                Conv $7\times7$  & 0.078/0.089  & 0.067/0.069 & 11.2M & 53.5G \\
                Conv $9\times9$  &  0.073/0.088 & 0.064/0.067 & 16.7M & 83.0G \\
                Conv $11\times11$  & 0.074/0.088 & 0.067/0.069 & 23.7M & 120G \\
                Conv $13\times13$  &  0.074/0.090 & 0.065/0.068 & 32.0M & 164G \\
                \hline
                ASPP \cite{pmnet} & 0.072/0.093  & 0.063/0.069 & 8.62M & 38.7G \\
                ASPP, $C=16$ & 0.088/0.097 &  0.070/0.073 & 2.2M & 9.72G \\
                ASPP, $d=2$ & 0.074/0.091 & 0.067/0.070 &  2.1M & 27.2G \\
                \hline
            \end{tabular}
            \caption{Comparison of training/validation loss and complexity in terms of the number of parameters and MACs for different variations of UNetDCN. 
                Results for the proposed model with no modification are listed in the first row for comparison.}
            \label{table:ablation}
        \end{table}

    \section{APPLICATION TO COVERAGE OPTIMIZATION}\label{sec:optimization}
        To showcase an application of the proposed models, we consider the problem of optimizing the directivity of cellular BS to 
        maximize the coverage in a given area. 
        We study two possible formulations of the problem that model different scenarios, utilizing the fast interference time and 
        differentiability of the CNN models.
        More precisely, we freeze the network after training and treat it as a
        differentiable function from the inputs, that include a representation of
        the antenna orientation which we aim to optimize, to the generated radio
        maps. Via backpropagation through the frozen network and gradient descent, we can optimize the antenna orientation parameters with respect to
        the coverage, that is defined as a function of the radio maps.

        \subsection{Coverage Definition and Optimization Objective}
            Assume that $M\in\mathbb{N}$ BS are installed at fixed positions on a given city map.
            The idealized antenna patterns considered are symmetric around the boresight, allowing us to characterize the orientation
            of a Tx by an azimuth angle $\varphi$ and an elevation angle $\theta$ with reference to the standard coordinate system. 
            Using any of the CNN models described before, we obtain an estimate of the path loss $\left(\PL_{i}(\varphi, \theta, x, y)\right)_{\dB}$ 
            for the $i$-th BS, with directivity defined by the azimuth angle $\varphi\in[0,2\pi]$ and elevation angle $\theta\in[0,\pi]$, 
            in any point $(x,y)\in\mathbb{T}$, where $\mathbb{T}\subset\mathbb{D}=\{1,\ldots,256\}\times\{1,\ldots,256\}$ is a given target area.
            Assuming a fixed input power $(P_T)_{\dBm}\in\mathbb{R}$ for all BS, we calculate the received power $P_i$ as a function 
            of $(\varphi, \theta, x, y)$ according to \eqref{eq:power_ratio} for each BS.

            As the first scenario, we consider a macro-diversity system \cite{macro-diversity}, in which signals from multiple BS are combined non-coherently and therefore the total received power in linear scale is defined as the sum
            \begin{equation}\label{eq:opt_power_tot}
                \left(\power_{tot}(\vec{\varphi}, \vec{\theta}, x, y)\right)_{\watts} = \sum_{i=1}^M \left(\power_i(\varphi_i, \theta_i, x, y)\right)_{\watts},
            \end{equation}
            with $\vec{\varphi}=(\varphi_i)_{i=1}^M, \vec{\theta}=(\theta_i)_{i=1}^M$.
            We aim to find $\vec{\varphi}\in[0,2\pi]^M, \vec{\theta}\in[0,\pi]^M$ that maximize the $10$-th percentile $p_{10}$ of \eqref{eq:opt_power_tot} in $\dBm$ scale
            over the locations in the target area, which is
            \begin{equation}\label{eq:opt_tot}
                \begin{split}
                    p_{10}(\vec{\varphi}, \vec{\theta}) = \min\{&t\in\mathbb{R}\,:\,(\power_{tot}(\vec{\varphi}, \vec{\theta}, x, y))_{\dBm}\leq t \\
                                                    &\text{ for } 10\% \text{ of the points }(x,y)\text{ in }\mathbb{T}\}.
                \end{split}
            \end{equation}
            In other words, we seek to maximize a lower bound for \eqref{eq:opt_power_tot} that holds for $90\%$ of the potential UE locations.
            Attempting to maximize the minimum value rather than the $10$-th percentile, which would be a lower bound for all locations, only worked for very restrictive target areas. 
            This is because some locations are difficult or impossible to cover by any Tx, and the models sometimes fail to accurately identify the exact edges of buildings at the pixel level.

            As a second scenario, we assume non-cooperative BS and aim to guarantee a strong signal from one of the BS while keeping the interference from the other BS low.
            More precisely, we consider the maximum of the signal-to-interference-plus-noise ratio (SINR) across all BS,
            \begin{equation}\label{eq:opt_sinr}
                \begin{split}
                    & \SINR(\vec{\varphi}, \vec{\theta}, x, y) \\
                    = &\max_{i=1,\ldots,M}\frac{\left(\power_i(\varphi_i, \theta_i, x, y)\right)_W}{\sum_{\substack{j\neq i}}\left(\power_j(\varphi_i, \theta_i, x, y)\right)_W + (\power_N)_W},
                \end{split}
            \end{equation}
            where $\power_N$ denotes the power of the noise in the system.
            This time, we aim to guarantee a sufficiently good SINR level in the largest possible area, i.e. find $\vec{\varphi}\in[0,2\pi]^M, \vec{\theta}\in[0,\pi]^M$ that maximize
            \begin{equation}\label{eq:opt_sinr_thresh}
                \# \{(x,y)\in\mathbb{T} \,:\, \SINR(\vec{\varphi}, \vec{\theta}, x, y) \geq t \}
            \end{equation}
            for some threshold $t>0$, where $\#$ denotes the cardinality of a set.

        \subsection{Implementation and Results}
        \begin{figure*}[!ht]
            \begin{subfigure}[t]{\textwidth}
                \centering
                \begin{minipage}{\textwidth}
                    \centering
                    \includegraphics[width=0.1\textwidth]{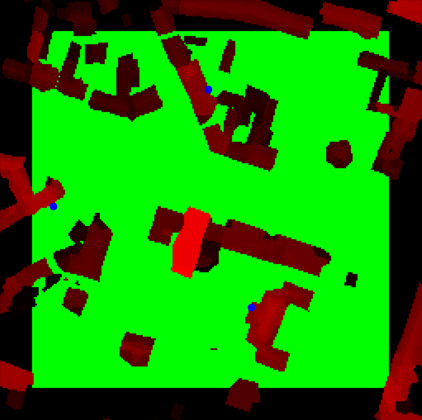}
                    \includegraphics[width=0.1\textwidth]{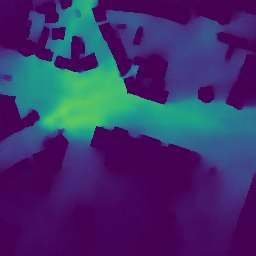}
                    \includegraphics[width=0.1\textwidth]{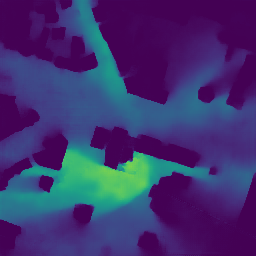}
                    \includegraphics[width=0.1\textwidth]{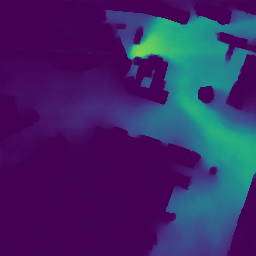}
                    \includegraphics[width=0.026\textwidth]{jaens5}
                    \raisebox{0.8cm}{$\overset{\eqref{eq:opt_power_tot}}{\mapsto}$}
                    \includegraphics[width=0.1\textwidth]{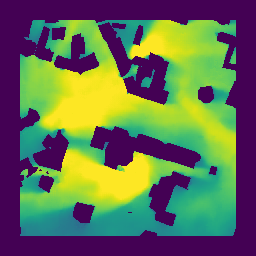}
                    \includegraphics[width=0.025\textwidth]{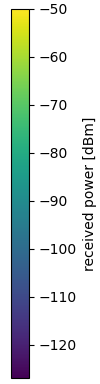}
                \end{minipage}
                \vspace{0.2cm}
                \\
                \begin{minipage}{\textwidth}
                    \centering
                    \includegraphics[width=0.1\textwidth]{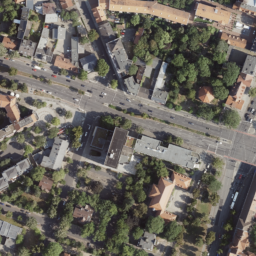}
                    \includegraphics[width=0.1\textwidth]{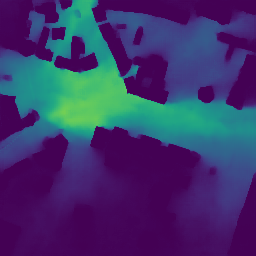}
                    \includegraphics[width=0.1\textwidth]{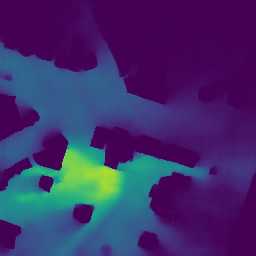}
                    \includegraphics[width=0.1\textwidth]{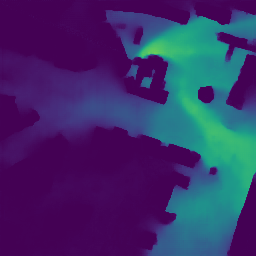}
                    \includegraphics[width=0.026\textwidth]{jaens5}
                    \raisebox{0.8cm}{$\overset{\eqref{eq:opt_power_tot}}{\mapsto}$}
                    \includegraphics[width=0.1\textwidth]{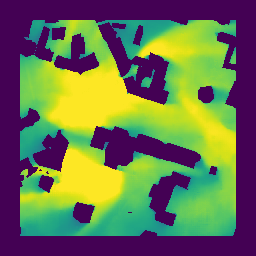}
                    \includegraphics[width=0.025\textwidth]{jaens109}
                \end{minipage}
                \caption{First scenario \eqref{eq:opt_tot}. 
                Top row: City map with buildings in red, target area in green and BS in blue, vegetation omitted, radio maps and coverage map with initial angles. 
                Second row: Aerial image, radio maps and coverage in the target area after optimization with gradient descent.}\label{fig:opt:scenario1}
            \end{subfigure}
            \vspace{0.4cm}
            \\
            \begin{subfigure}[t]{\textwidth}
                \begin{minipage}{\textwidth} 
                    \centering
                    \includegraphics[width=0.1\textwidth]{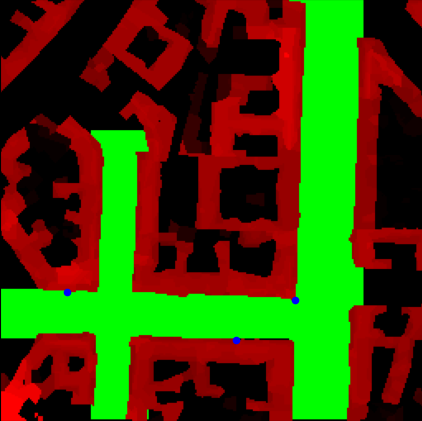}
                    \includegraphics[width=0.1\textwidth]{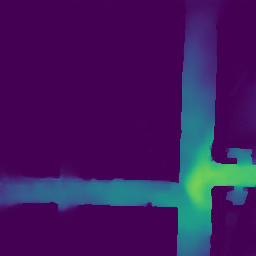}
                    \includegraphics[width=0.1\textwidth]{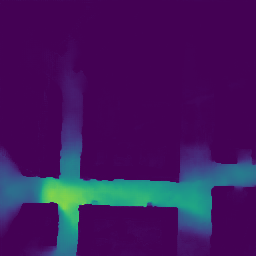}
                    \includegraphics[width=0.1\textwidth]{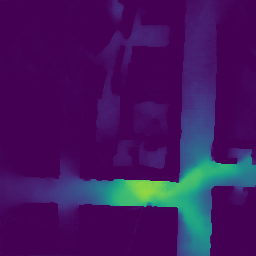}
                    \includegraphics[width=0.026\textwidth]{jaens5}
                    \raisebox{0.8cm}{$\overset{\eqref{eq:opt_sinr}}{\mapsto}$}
                    \includegraphics[width=0.1\textwidth]{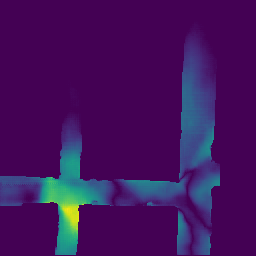}
                    \includegraphics[width=0.022\textwidth]{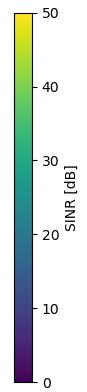}
                \end{minipage}
                \vspace{0.2cm}
                \\
                \begin{minipage}{\textwidth}
                    \centering
                    \includegraphics[width=0.1\textwidth]{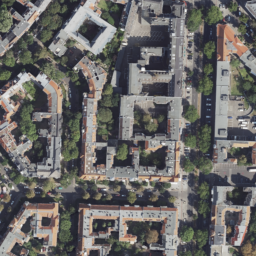}
                    \includegraphics[width=0.1\textwidth]{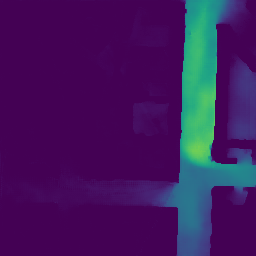}
                    \includegraphics[width=0.1\textwidth]{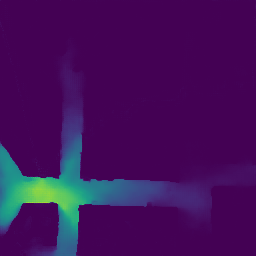}
                    \includegraphics[width=0.1\textwidth]{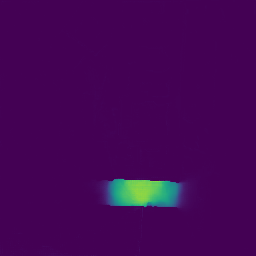}
                    \includegraphics[width=0.026\textwidth]{jaens5}
                    \raisebox{0.8cm}{$\overset{\eqref{eq:opt_sinr}}{\mapsto}$}
                \includegraphics[width=0.1\textwidth]{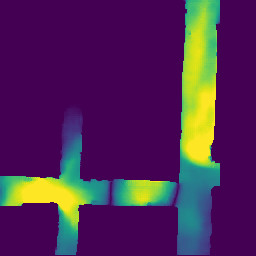}
                \includegraphics[width=0.022\textwidth]{jaens120}
            \end{minipage}
            \caption{Second scenario \eqref{eq:opt_sinr_thresh}.
            Top row: City map with buildings in red, target area in green and BS in blue, vegetation omitted, radio maps and coverage map with initial angles. 
            Second row: Aerial image, radio maps and coverage in the target area after optimization with gradient descent. }
            \label{fig:opt:scenario2}
        \end{subfigure}
        \caption{Coverage Optimization}
        \end{figure*}
            To predict the radio maps, we choose the DCN model that only receives an aerial image, the Tx location and the antenna gain as input.
            As starting and reference values for the optimization, we define the elevation angle between the $z$-axis (pointing straight up) and the antenna boresight as $\theta_0=\frac{3}{4}\pi$ 
            and we choose the azimuth angle $\varphi_0$ so that the boresight of the Tx points away from the building the BS is located on.
            The first optimization routine we consider is random search, i.e. trying angles $(\varphi,\theta)$ randomly drawn from $[\varphi_0-\frac{3}{4}\pi, \varphi_0+\frac{3}{4}\pi]\times[\pi/2, \pi]$ 
            for a fixed number of iterations.
            Note that we restrict the domain of the angles to avoid clearly suboptimal configurations in which the Tx points onto the building it is located on or towards the sky. 
            As a second option, we investigate leveraging the differentiability of the CNN model. 
            To achieve this, we define the angles as trainable parameters, freeze the network weights, and treat the antenna gain as a differentiable function of the angles.
            This allows us to optimize the angles of all BS simultaneously using gradient descent.
            Although it is not possible to calculate the true optima analytically or numerically, we can compare the final coverage score with that of the initial antenna configuration.

            \begin{table}[!htbp]
                \centering
                \begin{tabular}{|l|l|}
                    \hline
                    Initial angles & -86.5 \\
                    Random Search & -81.6 \\
                    Gradient Descent & -81.1 \\
                    \hline
                \end{tabular}        
                \caption{Results of scenario 1 \eqref{eq:opt_tot} in dBm}
                \label{table:opt:scenario1}
            \end{table}

            Minimizing \eqref{eq:opt_tot} directly with gradient descent is problematic, since in each iteration only the value at one or a few specific locations with the current $10$-th percentile
            is taken into account during backpropagation.
            To obtain an approximate formulation, we replace the $10$-th percentile in \eqref{eq:opt_tot} by the Boltzmann operator \cite{boltzmann}, defined as
            \begin{equation*}
                B_\alpha(S) = \frac{\sum_{x\in S} x\exp(\alpha x)}{\sum_{x\in S} \exp(\alpha x)},
            \end{equation*}
            for a finite set $S\subset\mathbb{R}$ and a parameter $\alpha<0$. 
            For negative values of $\alpha$ with large magnitude, $B_\alpha$ approximates the minimum and for $\alpha$ close to $0$ the mean of the set $S$.
            Similarly, the thresholding and counting in \eqref{eq:opt_sinr_thresh} are not differentiable.
            In our implementation, for gradient descent, we replace the thresholding operation by an appropriately shifted sigmoid function in order to obtain a smooth transition from $0$ to $1$
            for values below and above the threshold, and we implement the cardinality by summing over all pixels.
            \begin{table}[!htbp]
                \centering
                \begin{tabular}{|l|l|}
                    \hline
                    Initial angles & 3121 \\
                    Random Search & 11703 \\
                    Gradient Descent & 12338 \\
                    \hline
                \end{tabular}
                \caption{Results of scenario 2 \eqref{eq:opt_sinr_thresh}, number of pixels in the target area with SINR above the threshold}
                \label{table:opt:scenario2}
            \end{table}
            We consider $M=3$ BS, a $120^\circ$ sector antenna pattern, noise power of $-104$dBm (corresponding to a noise power spectral density of $-174$dBm/Hz and $10$MHz bandwidth), 
            and we assume a Tx input power of $23$dBm as detailed in Section \ref{sec:rm_simulations}.
            Both methods are run for $500$ iterations, which takes about $15$s with random search and $60$s with gradient descent, although we observed that gradient descent is already close 
            to the optimal value after about $50$ iterations.
            The SINR threshold in \eqref{eq:opt_sinr_thresh} is set to $t=20\dB$.

            The results are listed in Tables \ref{table:opt:scenario1} and \ref{table:opt:scenario2} and visualized in Figures \ref{fig:opt:scenario1} and \ref{fig:opt:scenario2}.
            Both methods provide a solid improvement over the baseline in scenario 1 and a large improvement in scenario 2, with gradient descent performing slightly better.
            This also shows that our approach of optimizing a differentiable approximation of the coverage score with gradient descent is indeed valid in order to maximize the true score.


    \section{CONCLUSION}\label{sec:conclusions}
        In this paper, we demonstrate that accurate path loss radio maps can be predicted even without full 3D environmental data, using only aerial images or with the addition of height but no classification information. 
        This opens up new possibilities, such as using data from a UAV flyover, where images and potentially a LiDAR scan are captured, allowing direct and efficient radio map prediction from this data. 

        As a further potential extension of this, predicting the radio map from satellite data available for the whole planet (similarly to \cite{plgan}, but in cellular networks) would allow large-scale applications to network planning and related tasks.
        The often lower spatial resolution may make the recognition of objects and shapes more difficult.
        On the other hand, incorporation of other spectral bands or radar data could be beneficial for the implicit classification of objects and height estimation \cite{sentinel_height}.

        In this work, we made certain assumptions, such as approximating the ground as flat, to simplify the modeling process. 
        We acknowledge that these simplifications may limit the applicability of our models to regions with significant elevation variations. 
        Future work could involve retraining models with datasets that include ground elevation data to better accommodate diverse geographic scenarios, as in \cite{fadenet}.
        While we have presented the radiomap prediction results in terms of the RMSE averaged over the whole map,
        in line with the existing literature, a more refined study of the ``conditional'' RMSE (e.g., RMSE versus distance from Tx, or RMSE conditioned on
        (non-)line-of-sight propagation) is certainly an interesting topic for future work.

        Although this study does not cover it, the dataset could be used to explore joint semantic segmentation and height estimation from aerial images \cite{im2height}, \cite{joint_seg_height}. 
        This could serve as an intermediate step toward improving radio map predictions.

        By making our dataset and code available for public use, we aim to facilitate the work of other researchers and promote reproducible and comparable investigations.
        Featuring directive Tx antennas, the dataset opens potential for the investigation of more downstream tasks in 5G/6G networks, such as beam codebook design and beam management \cite{beams}.

    \section*{ACKNOWLEDGMENT}
        We thank \c{C}a\v{g}kan Yapar for advice on the dataset generation and the related literature, Ron Levie for discussions on how to generate and use line-of-sight information, Tom Burgert for advice on curriculum learning and the recommendation to use PyTorch Lightning and Saeid Dehkordi for proofreading.

    \bibliographystyle{IEEEtran}
    \bibliography{bib_rmdir}

\begin{thebibliography}{10}
\providecommand{\url}[1]{#1}
\csname url@samestyle\endcsname
\providecommand{\newblock}{\relax}
\providecommand{\bibinfo}[2]{#2}
\providecommand{\BIBentrySTDinterwordspacing}{\spaceskip=0pt\relax}
\providecommand{\BIBentryALTinterwordstretchfactor}{4}
\providecommand{\BIBentryALTinterwordspacing}{\spaceskip=\fontdimen2\font plus
\BIBentryALTinterwordstretchfactor\fontdimen3\font minus
  \fontdimen4\font\relax}
\providecommand{\BIBforeignlanguage}[2]{{%
\expandafter\ifx\csname l@#1\endcsname\relax
\typeout{** WARNING: IEEEtran.bst: No hyphenation pattern has been}%
\typeout{** loaded for the language `#1'. Using the pattern for}%
\typeout{** the default language instead.}%
\else
\language=\csname l@#1\endcsname
\fi
#2}}
\providecommand{\BIBdecl}{\relax}
\BIBdecl

\bibitem{link_scheduling}
X.~Wu, S.~Tavildar, S.~Shakkottai, T.~Richardson, J.~Li, R.~Laroia, and
  A.~Jovicic, ``Flashlinq: A synchronous distributed scheduler for peer-to-peer
  ad hoc networks,'' \emph{IEEE/ACM Trans. Netw.}, vol.~21, no.~4, pp.
  1215--1228, 8 2013.

\bibitem{locunet}
C.~Yapar, R.~Levie, G.~Kutyniok, and G.~Caire, ``Real-time outdoor localization
  using radio maps: A deep learning approach,'' \emph{IEEE Transactions on
  Wireless Communications}, vol.~22, no.~12, pp. 9703--9717, 2023.

\bibitem{beams}
D.~E. Berraki, S.~M.~D. Armour, and A.~R. Nix, ``Codebook based beamforming and
  multiuser scheduling scheme for mmwave outdoor cellular systems in the 28, 38
  and 60ghz bands,'' in \emph{2014 IEEE Globecom Workshops}.\hskip 1em plus
  0.5em minus 0.4em\relax IEEE, 12 2014, pp. 382--387.

\bibitem{ray-tracing-accuracy}
M.~Lecci, P.~Testolina, M.~Polese, M.~Giordani, and M.~Zorzi, ``Accuracy versus
  complexity for mmwave ray-tracing: A full stack perspective,'' \emph{IEEE
  Trans. Wireless Commun.}, vol.~20, no.~12, pp. 7826--7841, 12 2021.

\bibitem{pmnet}
J.-H. Lee and A.~F. Molisch, ``A scalable and generalizable pathloss map
  prediction,'' \emph{IEEE Transactions on Wireless Communications}, vol.~23,
  no.~11, pp. 17\,793--17\,806, 2024.

\bibitem{applications}
\BIBentryALTinterwordspacing
S.~Bakirtzis, C.~Yapar, M.~Fiore, J.~Zhang, and I.~Wassell, ``Empowering
  wireless network applications with deep learning-based radio propagation
  models,'' 2024. [Online]. Available: \url{https://arxiv.org/abs/2408.12193}
\BIBentrySTDinterwordspacing

\bibitem{3gpp_log_normal}
G.~Calcev, D.~Chizhik, B.~Goransson, S.~Howard, H.~Huang, A.~Kogiantis, A.~F.
  Molisch, A.~L. Moustakas, D.~Reed, and H.~Xu, ``A wideband spatial channel
  model for system-wide simulations,'' \emph{IEEE Trans. Veh. Technol.},
  vol.~56, no.~2, pp. 389--403, 3 2007.

\bibitem{ray-tracing}
Z.~Yun and M.~F. Iskander, ``Ray tracing for radio propagation modeling:
  Principles and applications,'' \emph{IEEE Access}, vol.~3, pp. 1089--1100,
  2015.

\bibitem{rm_completion}
M.~Kasparick, R.~L.~G. Cavalcante, S.~Valentin, S.~Stańczak, and M.~Yukawa,
  ``Kernel-based adaptive online reconstruction of coverage maps with side
  information,'' \emph{IEEE Trans. Veh. Technol.}, vol.~65, no.~7, pp.
  5461--5473, 7 2016.

\bibitem{ml_single}
S.~I. Popoola, A.~Jefia, A.~A. Atayero, O.~Kingsley, N.~Faruk, O.~F. Oseni, and
  R.~O. Abolade, ``Determination of neural network parameters for path loss
  prediction in very high frequency wireless channel,'' \emph{IEEE Access},
  vol.~7, pp. 150\,462--150\,483, 2019.

\bibitem{cnn}
Z.~Li, F.~Liu, W.~Yang, S.~Peng, and J.~Zhou, ``A survey of convolutional
  neural networks: Analysis, applications, and prospects,'' \emph{IEEE Trans.
  Neural Netw. Learn. Syst.}, vol.~33, no.~12, pp. 6999--7019, 12 2022.

\bibitem{radiounet}
R.~Levie, c.~Yapar, G.~Kutyniok, and G.~Caire, ``Radiounet: Fast radio map
  estimation with convolutional neural networks,'' \emph{IEEE Trans. Wireless
  Commun.}, vol.~20, no.~6, pp. 4001--4015, 6 2021.

\bibitem{unet}
O.~Ronneberger, P.~Fischer, and T.~Brox, ``U-net: Convolutional networks for
  biomedical image segmentation,'' in \emph{Medical Image Computing and
  Computer-Assisted Intervention -- MICCAI 2015}, N.~Navab, J.~Hornegger, W.~M.
  Wells, and A.~F. Frangi, Eds.\hskip 1em plus 0.5em minus 0.4em\relax Cham:
  Springer International Publishing, 2015, pp. 234--241.

\bibitem{plnet}
X.~Zhang, X.~Shu, B.~Zhang, J.~Ren, L.~Zhou, and X.~Chen, ``Cellular network
  radio propagation modeling with deep convolutional neural networks,'' in
  \emph{Proceedings of the 26th ACM SIGKDD International Conference on
  Knowledge Discovery \& Data Mining}, ser. KDD '20.\hskip 1em plus 0.5em minus
  0.4em\relax Association for Computing Machinery, 8 2020, pp. 2378--2386.

\bibitem{fadenet}
V.~V. Ratnam, H.~Chen, S.~Pawar, B.~Zhang, C.~J. Zhang, Y.-J. Kim, S.~Lee,
  M.~Cho, and S.-R. Yoon, ``Fadenet: Deep learning-based mm-wave large-scale
  channel fading prediction and its applications,'' \emph{IEEE Access}, vol.~9,
  pp. 3278--3290, 2021.

\bibitem{radiotrans}
Y.~Tian, S.~Yuan, W.~Chen, and N.~Liu, ``Transformer based radio map prediction
  model for dense urban environments,'' in \emph{2021 13th International
  Symposium on Antennas, Propagation and EM Theory (ISAPE)}, vol.~1, 2021, pp.
  1--3.

\bibitem{dilated}
F.~Yu and V.~Koltun, ``Multi-scale context aggregation by dilated
  convolutions,'' in \emph{International Conference on Learning
  Representations}, 2016.

\bibitem{ziemann21}
M.~R. Ziemann, J.~S. Hyatt, and M.~S. Lee, ``Convolutional neural networks for
  radio frequency ray tracing,'' in \emph{MILCOM 2021 - 2021 IEEE Military
  Communications Conference (MILCOM)}.\hskip 1em plus 0.5em minus 0.4em\relax
  IEEE, 11 2021, pp. 618--622.

\bibitem{deepray}
S.~Bakirtzis, K.~Qiu, J.~Zhang, and I.~Wassell, ``Deepray: Deep learning meets
  ray-tracing,'' in \emph{2022 16th European Conference on Antennas and
  Propagation (EuCAP)}, 2022, pp. 1--5.

\bibitem{vit}
\BIBentryALTinterwordspacing
A.~Dosovitskiy, L.~Beyer, A.~Kolesnikov, D.~Weissenborn, X.~Zhai,
  T.~Unterthiner, M.~Dehghani, M.~Minderer, G.~Heigold, S.~Gelly, J.~Uszkoreit,
  and N.~Houlsby, ``An image is worth 16x16 words: Transformers for image
  recognition at scale,'' in \emph{International Conference on Learning
  Representations}, 2021. [Online]. Available:
  \url{https://openreview.net/forum?id=YicbFdNTTy}
\BIBentrySTDinterwordspacing

\bibitem{qiu22}
K.~Qiu, S.~Bakirtzis, H.~Song, J.~Zhang, and I.~Wassell, ``Pseudo ray-tracing:
  Deep leaning assisted outdoor mm-wave path loss prediction,'' \emph{IEEE
  Wireless Commun. Lett.}, vol.~11, no.~8, pp. 1699--1702, 8 2022.

\bibitem{plgan}
A.~Marey, M.~Bal, H.~F. Ates, and B.~K. Gunturk, ``Pl-gan: Path loss prediction
  using generative adversarial networks,'' \emph{IEEE Access}, vol.~10, pp.
  90\,474--90\,480, 2022.

\bibitem{cagkan3d}
C.~Yapar, R.~Levie, G.~Kutyniok, and G.~Caire. (2022) Dataset of pathloss and
  toa radio maps with localization application.

\bibitem{jaensch2024radiomapestimation}
\BIBentryALTinterwordspacing
F.~Jaensch, G.~Caire, and B.~Demir, ``Radio map estimation -- an open dataset
  with directive transmitter antennas and initial experiments,'' 2024.
  [Online]. Available: \url{https://arxiv.org/abs/2402.00878}
\BIBentrySTDinterwordspacing

\bibitem{dcn}
J.~Dai, H.~Qi, Y.~Xiong, Y.~Li, G.~Zhang, H.~Hu, and Y.~Wei, ``Deformable
  convolutional networks,'' in \emph{IEEE International Conference on Computer
  Vision (ICCV)}.\hskip 1em plus 0.5em minus 0.4em\relax IEEE, 10 2017, pp.
  764--773.

\bibitem{nanophotonics_inverse_design}
\BIBentryALTinterwordspacing
J.~Peurifoy, Y.~Shen, L.~Jing, Y.~Yang, F.~Cano-Renteria, B.~G. DeLacy, J.~D.
  Joannopoulos, M.~Tegmark, and M.~Solja{\v c}i{\'c}, ``Nanophotonic particle
  simulation and inverse design using artificial neural networks,''
  \emph{Science Advances}, vol.~4, no.~6, p. eaar4206, 2018. [Online].
  Available: \url{https://www.science.org/doi/abs/10.1126/sciadv.aar4206}
\BIBentrySTDinterwordspacing

\bibitem{sionna}
J.~Hoydis, S.~Cammerer, F.~{A\"{i}t Aoudia}, A.~Vem, N.~Binder, G.~Marcus, and
  A.~Keller, ``Sionna: An open-source library for next-generation physical
  layer research,'' 3 2022.

\bibitem{sionna_rt}
J.~Hoydis, F.~A. Aoudia, S.~Cammerer, M.~Nimier-David, N.~Binder, G.~Marcus,
  and A.~Keller, ``Sionna rt: Differentiable ray tracing for radio propagation
  modeling,'' 2023.

\bibitem{molisch}
A.~F. Molisch, \emph{Wireless Communications}, 2nd~ed.\hskip 1em plus 0.5em
  minus 0.4em\relax Wiley Publishing, 2011.

\bibitem{wssus}
P.~Bello, ``Characterization of randomly time-variant linear channels,''
  \emph{IEEE Transactions on Communications Systems}, vol.~11, no.~4, pp.
  360--393, 1963.

\bibitem{proakis}
J.~G. Proakis and M.~Salehi, \emph{\BIBforeignlanguage{eng}{Digital
  communications}}, 5th~ed.\hskip 1em plus 0.5em minus 0.4em\relax Boston u.a.:
  McGraw-Hill, 2008.

\bibitem{geoportal_berlin}
\BIBentryALTinterwordspacing
B.~u.~W. Senatsverwaltung~für Stadtentwicklung. Geoportal berlin. [Online].
  Available: \url{https://fbinter.stadt-berlin.de/fb/index.jsp}
\BIBentrySTDinterwordspacing

\bibitem{lastools}
\BIBentryALTinterwordspacing
rapidlasso GmbH. Lastools - efficient lidar processing software. [Online].
  Available: \url{http://rapidlasso.com/LAStools}
\BIBentrySTDinterwordspacing

\bibitem{wi}
\BIBentryALTinterwordspacing
Remcom. Wireless insite. [Online]. Available:
  \url{https://www.remcom.com/wireless-insite-em-propagation-software/}
\BIBentrySTDinterwordspacing

\bibitem{geoportal_bb}
\BIBentryALTinterwordspacing
L.~und Geobasisinformation~Brandenburg. Geobroker. [Online]. Available:
  \url{https://geobroker.geobasis-bb.de/}
\BIBentrySTDinterwordspacing

\bibitem{challenge}
c.~Yapar, F.~Jaensch, R.~Levie, G.~Kutyniok, and G.~Caire, ``The first pathloss
  radio map prediction challenge,'' in \emph{ICASSP 2023 - 2023 IEEE
  International Conference on Acoustics, Speech and Signal Processing
  (ICASSP)}, 2023, pp. 1--2.

\bibitem{deepspeed}
\BIBentryALTinterwordspacing
Microsoft. Deepspeed flops profiler. [Online]. Available:
  \url{https://www.deepspeed.ai}
\BIBentrySTDinterwordspacing

\bibitem{im2height}
S.~Du, J.~Xing, S.~Du, X.~Cui, X.~Xiao, W.~Li, and S.~Wang, ``Img2height:
  height estimation from single remote sensing image using a deep convolutional
  encoder-decoder network,'' \emph{International Journal of Remote Sensing},
  vol.~44, no.~18, pp. 5686--5712, 9 2023.

\bibitem{macro-diversity}
F.~Kirsten, D.~Öhmann, M.~Simsek, and G.~P. Fettweis, ``On the utility of
  macro- and microdiversity for achieving high availability in wireless
  networks,'' in \emph{2015 IEEE 26th Annual International Symposium on
  Personal, Indoor, and Mobile Radio Communications (PIMRC)}, 2015, pp.
  1723--1728.

\bibitem{boltzmann}
\BIBentryALTinterwordspacing
K.~Asadi and M.~L. Littman, ``An alternative softmax operator for reinforcement
  learning,'' in \emph{Proceedings of the 34th International Conference on
  Machine Learning}, ser. Proceedings of Machine Learning Research, D.~Precup
  and Y.~W. Teh, Eds., vol.~70.\hskip 1em plus 0.5em minus 0.4em\relax PMLR, 8
  2017, pp. 243--252. [Online]. Available:
  \url{https://proceedings.mlr.press/v70/asadi17a.html}
\BIBentrySTDinterwordspacing

\bibitem{sentinel_height}
D.~Frantz, F.~Schug, A.~Okujeni, C.~Navacchi, W.~Wagner, S.~{van der Linden},
  and P.~Hostert, ``National-scale mapping of building height using sentinel-1
  and sentinel-2 time series,'' \emph{Remote Sensing of Environment}, vol. 252,
  p. 112128, 2021.

\bibitem{joint_seg_height}
S.~Srivastava, M.~Volpi, and D.~Tuia, ``Joint height estimation and semantic
  labeling of monocular aerial images with cnns,'' in \emph{2017 IEEE
  International Geoscience and Remote Sensing Symposium (IGARSS)}, 2017, pp.
  5173--5176.

\end{thebibliography}

    \begin{IEEEbiography}[{\includegraphics[width=1in]{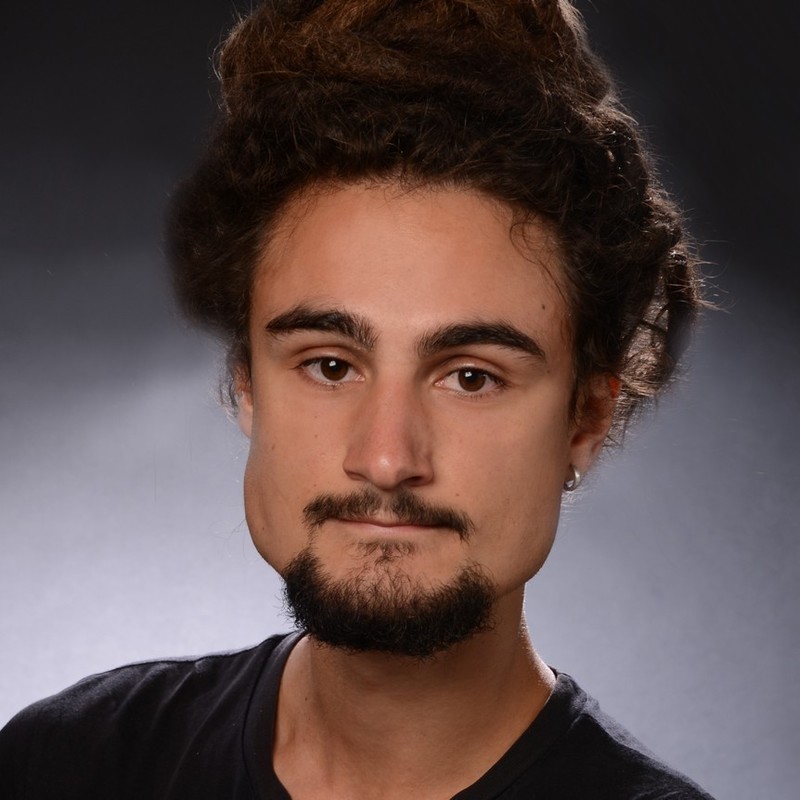}}]%
        {Fabian Jaensch} 
        received the B.Sc. and M.Sc. in Mathematics from Technische Universität Berlin in 2016 and 2021, respectively.
        He is currently working as a Ph.D. student at the Communications and Information Theory Chair, Technische Universität Berlin.
        His research interests include machine learning, in particular computer vision, and applications to wireless communications.

    \end{IEEEbiography}

    \begin{IEEEbiography}[{\includegraphics[width=1in]{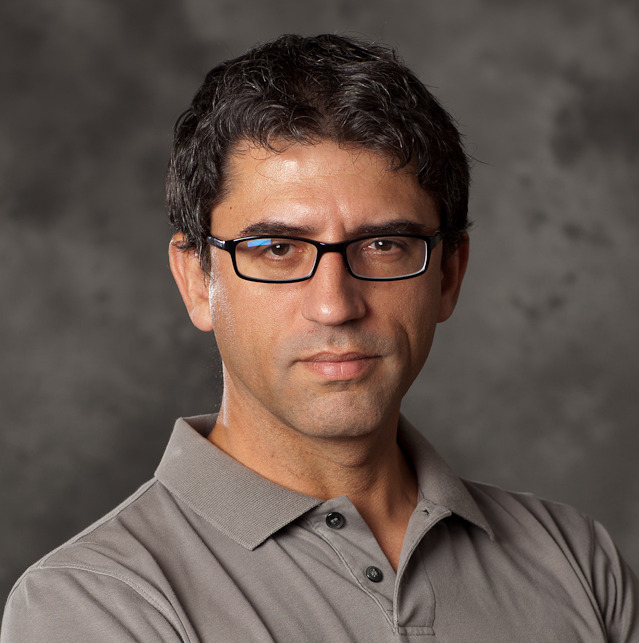}}]%
        {Giuseppe Caire} (S '92 -- M '94 -- SM '03 -- F '05) 
        was born in Torino in 1965. He received a
        B.Sc. in Electrical Engineering  from Politecnico di Torino in 1990, 
        an M.Sc. in Electrical Engineering from Princeton University in 1992, and a Ph.D. from Politecnico di Torino in 1994. 
        He has been a post-doctoral research fellow with the European Space Agency (ESTEC, Noordwijk, The Netherlands) in 1994-1995,
        Assistant Professor in Telecommunications at the Politecnico di Torino, Associate Professor at the University of Parma, Italy, 
        Professor with the Department of Mobile Communications at the Eurecom Institute,  Sophia-Antipolis, France,
        a Professor of Electrical Engineering with the Viterbi School of Engineering, University of Southern California, Los Angeles,
        and he is currently an Alexander von Humboldt Professor with the Faculty of Electrical Engineering and Computer Science at the Technical University of Berlin, Germany.

        He received the Jack Neubauer Best System Paper Award from the IEEE Vehicular Technology Society in 2003,  the
        IEEE Communications Society and Information Theory Society Joint Paper Award in 2004, in 2011, and in 2025, 
        the Okawa Research Award in 2006,   the Alexander von Humboldt Professorship in 2014, the Vodafone Innovation Prize in 2015, an ERC Advanced Grant in 2018,  
        the Leonard G. Abraham Prize for best IEEE JSAC paper in 2019, the IEEE Communications Society Edwin Howard Armstrong Achievement Award in 2020, the 2021 Leibniz Prize  
        of the German National Science Foundation (DFG), and the  CTTC Technical Achievement Award of the IEEE Communications Society in 2023.  Giuseppe Caire is a Fellow of IEEE since 2005.  
        He has served in the Board of Governors of the IEEE Information Theory Society from 2004 to 2007, and as officer from 2008 to 2013. He was President of the IEEE 
        Information Theory Society in 2011. 
        His main research interests are in the field of communications theory, information theory, channel and source coding
        with particular focus on wireless communications.   

    \end{IEEEbiography}

    \begin{IEEEbiography}[{\includegraphics[width=1in]{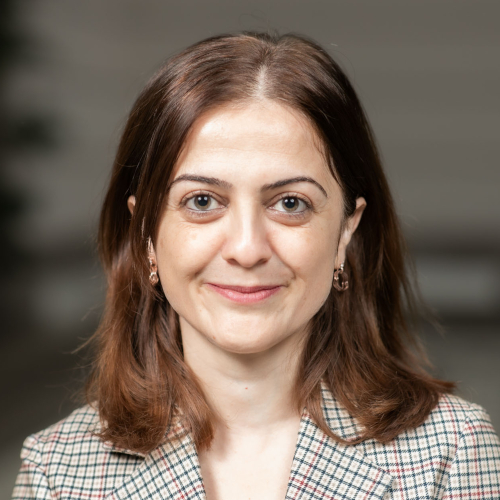}}]%
        {Begüm Demir} (S'06-M'11-SM'16) received the B.Sc., M.Sc., and Ph.D. degrees in electronic and telecommunication engineering from Kocaeli University, Kocaeli, Turkey, in 2005, 2007, 
        and 2010, respectively.

        She is currently a Full Professor and the founder head of the Remote Sensing Image Analysis (RSiM) group at the Faculty of Electrical Engineering and Computer Science, TU Berlin and 
        the head of the Big Data Analytics for Earth Observation research group at the Berlin Institute for the Foundations of Learning and Data (BIFOLD). Her research activities lie at the 
        intersection of machine learning, remote sensing and signal processing. Specifically, she performs research in the field of processing and analysis of large-scale Earth observation 
        data acquired by airborne and satellite-borne systems. She was awarded by the prestigious ‘2018 Early Career Award’ by the IEEE Geoscience and Remote Sensing Society for her research 
        contributions in machine learning for information retrieval in remote sensing. In 2018, she received a Starting Grant from the European Research Council (ERC) for her project 
        “BigEarth: Accurate and Scalable Processing of Big Data in Earth Observation”. She is an IEEE Senior Member and Fellow of European Lab for Learning and Intelligent Systems (ELLIS).

        Dr. Demir is a Scientific Committee member of several international conferences and workshops. She is a referee for several journals such as the Proceedings of the IEEE, the IEEE 
        Transactions on Geoscience and Remote Sensing, the IEEE Geoscience and Remote Sensing Letters, the IEEE Transactions on Image Processing, Pattern Recognition, the IEEE Transactions 
        on Circuits and Systems For Video Technology, the IEEE Journal of Selected Topics in Signal Processing, the International Journal of Remote Sensing, and several international 
        conferences. Currently she is an Associate Editor for the IEEE Geoscience and Remote Sensing Magazine.

    \end{IEEEbiography}
    
\end{document}